\documentclass[]{IEEEtran}

\usepackage{amsmath}
\usepackage{amsfonts}
\usepackage{amssymb}
\usepackage{amsthm}
\usepackage{graphics}
\usepackage{graphicx}
\usepackage{color}
\usepackage{tensor}
\usepackage[flushleft]{threeparttable}
\usepackage{makecell}
\usepackage{mathrsfs}
\usepackage{textcomp} 
\usepackage{multirow}
\usepackage{xcolor,colortbl}
\usepackage{longtable}
\usepackage{supertabular}
\usepackage{acro}

\newtheorem{theorem}{Theorem}[section]

\theoremstyle{definition}
\newtheorem{definition}{Definition}[section]

\theoremstyle{remark}
\newtheorem{remark}{Remark}

\DeclareAcronym{UE}{short=UE,long=User Equipment,}
\DeclareAcronym{RAN}{short=RAN,long=Radio Access Network,}
\DeclareAcronym{URLLC}{short=URLLC,long=Ultra-Reliable and
Low-Latency Communication,}
\DeclareAcronym{eMBB}{short=eMBB,long=enhanced Mobile BroadBand,}
\DeclareAcronym{VNF}{short=VNF,long=Virtualized Network Function,}
\DeclareAcronym{RU}{short=RU,long=Radio Unit,}
\DeclareAcronym{DU}{short=DU,long=Distributed Unit,}
\DeclareAcronym{CU}{short=CU,long=Central Unit,}
\DeclareAcronym{MNO}{short=MNO,long=Mobile Network Operator,}

\DeclareAcronym{AR}{short=AR,long=Augmented Reality,}
\DeclareAcronym{VR}{short=VR,long=Virtual Reality,}
\DeclareAcronym{UAV}{short=UAV,long=Unmanned Aerial Vehicle,}
\DeclareAcronym{NGMN}{short=NGMN,long=Next Generation Mobile Network,}
\DeclareAcronym{SMO}{short=SMO,long=Service Management and Orchestration
,}
\DeclareAcronym{TN}{short=TN,long=Transport Network,}
\DeclareAcronym{CN}{short=CN,long=Core Network,}
\DeclareAcronym{NSMF}{short=NSMF,long=Network
Slice Management Function,}
\DeclareAcronym{NSI}{short=NSI,long=Network Slice Instance,}
\DeclareAcronym{NSSI}{short=NSSI,long=Network Slice Subnet Instance,}
\DeclareAcronym{PDU}{short=PDU,long=Packet Data Unit,}
\DeclareAcronym{PLMN-ID}{short=PLMN-ID,long=Public Land Mobile Network IDentifier,}
\DeclareAcronym{S-NSSAI}{short=S-NSSAI,long=Single-Network Slice Selection Assistance Information,}
\DeclareAcronym{SST}{short=SST,long=Slice Service Type,}
\DeclareAcronym{SD}{short=SD,long=Slice Differentiator,}
\DeclareAcronym{mMTC}{short=mMTC,long=massive Machine Type Communications,}
\DeclareAcronym{B5G}{short=B5G,long=Beyond 5G,}
\DeclareAcronym{QoS}{short=QoS,long=Quality of Service,}
\DeclareAcronym{SDN}{short=SDN,long=Software Defined Network,}
\DeclareAcronym{OMNeT++}{short=OMNeT++,long=Objective Modular Network Testbed in C++,}
\DeclareAcronym{DiscoDNC}{short=DiscoDNC,long=Disco Deterministic Network Calculator,}
\DeclareAcronym{NG-RAN}{short=NG-RAN,long=Next Generation Radio Access Network,}
\DeclareAcronym{5GC}{short=5GC,long=5G Core,}
\DeclareAcronym{LTE}{short=LTE,long=Long Term Evolution,}
\DeclareAcronym{E-UTRA}{short=E-UTRA,long=Evolved Universal Terrestrial Radio Access
,}
\DeclareAcronym{LLS}{short=LLS,long=Lower Layer Split,}
\DeclareAcronym{HLS}{short=HLS,long=Higher Layer Split,}
\DeclareAcronym{RF}{short=RF,long=Radio Frequency,}
\DeclareAcronym{vCU}{short=vCU,long=virtualized CU,}
\DeclareAcronym{vDU}{short=vDU,long=virtualized DU,}
\DeclareAcronym{SCF}{short=SCF,long=Small Cell Forum,}
\DeclareAcronym{eCPRI}{short=eCPRI,long=enhanced Common Public Radio Interface,}
\DeclareAcronym{FIFO}{short=FIFO,long=First-In-First-Out,}
\DeclareAcronym{V2X}{short=V2X,long=Vehicle-to-Everything,}
\DeclareAcronym{SLA}{short=SLA,long=Service Level Agreement,}
\DeclareAcronym{IoT}{short=IoT,long=Internet-of-Things,}
\DeclareAcronym{DNC}{short=DNC,long=Deterministic Network Calculus,}
\DeclareAcronym{PBOO}{short=PBOO,long=Pay Bursts Only Once,}
\DeclareAcronym{ICMP}{short=ICMP,long=Internet Control Message Protocol,}
\DeclareAcronym{MTU}{short=MTU,long=Maximum Transmission Unit,}
\DeclareAcronym{CD}{short=CD,long=Centralization Degree,}
\DeclareAcronym{CR}{short=CR,long=Computing Resources,}
\DeclareAcronym{MEC}{short=MEC,long=Multi-Access Edge Computing,}
\DeclareAcronym{DRC}{short=DRC,long=Disaggregated RAN Combination,}
\DeclareAcronym{ML}{short=ML,long=Machine Learning,}
\DeclareAcronym{AI}{short=AI,long=Artificial Intelligence,}
\DeclareAcronym{SNC}{short=SNC,long=Stochastic Network Calculus,}
\DeclareAcronym{QSI}{short=QSI,long=Queue State Information,}
\DeclareAcronym{3GPP}{short=3GPP,long=3rd Generation Partnership Project,}
\DeclareAcronym{PSO}{short=PSO,long=Particle Swarm Optimization,}
\DeclareAcronym{NR}{short=NR,long=New Radio}
\DeclareAcronym{ARPU}{short=ARPU,long=Average Revenue Per User}
\DeclareAcronym{SFA}{short=SFA,long=Single Flow Analysis}
\DeclareAcronym{TFA}{short=TFA,long=Total Flow Analysis}
\DeclareAcronym{WRR}{short=WRR,long=Weighted Round Robin}
\DeclareAcronym{RB}{short=RB,long=Resource Block}
\DeclareAcronym{MCS}{short=MCS,long=Modulation and Coding Scheme}
\DeclareAcronym{GPS}{short=GPS,long=Generalized Processor Sharing}
\DeclareAcronym{gNB}{short=gNB,long=Next Generation Node B}
\DeclareAcronym{ng-eNB}{short=ng-eNB,long=Next Generation e-Node B}
\DeclareAcronym{D-RAN}{short=D-RAN,long=Distributed-RAN}
\DeclareAcronym{BE}{short=BE,long=Break-Even}
\DeclareAcronym{NC}{short=NC,long=Network Calculus}
\DeclareAcronym{PD}{short=PD,long=Propagation Delay}
\DeclareAcronym{FoI}{short=FoI,long=Flow of Interest}
\DeclareAcronym{RC}{short=RC,long=Reference Core}
\DeclareAcronym{FFT}{short=FFT,long=Fast Fourier Transform}
\DeclareAcronym{KPI}{short=KPI,long=Key Performance Indicator}
\DeclareAcronym{TTI}{short=TTI,long=Transmission Time Interval}
\DeclareAcronym{TBS}{short=TTI,long=Transport Block Size}
\DeclareAcronym{PDSCH}{short=PDSCH,long=Physical Data Shared Channel}
\DeclareAcronym{RRC}{short=RRC,long=Radio Resource Control}
\DeclareAcronym{PDCP}{short=PDCP,long=Packet Data Convergence Protocol}
\DeclareAcronym{RLC}{short=RLC,long=Radio Link Control}
\DeclareAcronym{MAC}{short=MAC,long=Medium Access Control}
\DeclareAcronym{FAPI}{short=FAPI,long=Functional Application Platform Interface}
\DeclareAcronym{CFI}{short=CFI,long=Control Format Indicator}
\DeclareAcronym{PHY}{short=PHY,long=Physical Layer}
\DeclareAcronym{ARB}{short=ARB,long=Arbitrary multiplexing}
\DeclareAcronym{PNF}{short=PNF,long=Physical Network Function}
\DeclareAcronym{PMOO}{short=PMOO,long=Pay Multiplexing Only Once}
\DeclareAcronym{FFS}{short=FFS,long=Flexible Functional Split}

\usepackage[symbol]{footmisc}

\definecolor{LightCyan}{rgb}{0.88,1,1}
\definecolor{Gray}{gray}{0.90}
\definecolor{Gray1}{gray}{0.60}

\title{}
\author{Flávio G. C. Rocha, 
Gabriel M. F. de Almeida, 
Kleber V. Cardoso,
Cristiano B. Both, 
and José F. de Rezende
  \IEEEcompsocitemizethanks{
    \IEEEcompsocthanksitem  Flávio G. C. Rocha, Gabriel M. F. de Almeida and Kleber V. Cardoso are with Universidade Federal de Goi\'{a}s (UFG), Goi\^{a}nia, Brazil: flaviogcr@ufg.br, \{gabrielmatheus, kleber\}@inf.ufg.br; Cristiano B. Both is with Universidade do Vale do Rio dos Sinos (UNISINOS), Porto Alegre, Brazil:~cbboth@unisinos.br; José F. de Rezende is with Universidade Federal do Rio de Janeiro (UFRJ), Rio de Janeiro, Brazil: rezende@land.ufrj.br.
  }
  }



\makeatletter
\newcommand*\titleheader[1]{\gdef\@titleheader{#1}}
\AtBeginDocument{%
  \let\st@red@title\@title%
  \def\@title{%
    \bgroup\normalfont\large\raggedright\@titleheader\par\egroup
    \vskip1.5em\st@red@title}
}
\makeatother

\titleheader{2017 IEEE 999999th International Something Conference}

\begin{document}

\onecolumn 
\pagestyle{empty} 
\begin{center}
  \large\bfseries  
Notice: This work has been submitted to the IEEE for possible publication. Copyright may be transferred without notice, after which this version may no longer be accessible.
\end{center}
\twocolumn 
\setcounter{page}{1} 

\title{Optimal Resource Allocation with Delay Guarantees for Network Slicing in Disaggregated RAN}


\maketitle

\begin{abstract}
In this article, we propose a novel formulation for the resource allocation problem of a sliced and disaggregated Radio Access Network (RAN) and its transport network. Our proposal assures an end-to-end delay bound for the \ac{URLLC} use case while jointly considering the number of admitted users, the transmission rate allocation per slice, the functional split of RAN nodes and the routing paths in the transport network. We use deterministic network calculus theory to calculate delay along the transport network connecting disaggregated RANs deploying network functions at the \ac{RU}, \ac{DU}, and \ac{CU} nodes. The maximum end-to-end delay is a constraint in the optimization-based formulation that aims to maximize \ac{MNO} profit, considering a cash flow analysis to model revenue and operational costs using data from one of the world's leading \acp{MNO}. The optimization model leverages a \ac{FFS} approach to provide a new degree of freedom to the resource allocation strategy. Simulation results reveal that, due to its non-linear nature, there is no trivial solution to the proposed optimization problem formulation. Our proposal guarantees a maximum delay for \ac{URLLC} services while satisfying minimal bandwidth requirements for \ac{eMBB} services and maximizing the \ac{MNO}'s profit. 

\vspace{3mm}

\end{abstract} 
\section{Introduction}\label{sec:introduction}

Fifth-generation mobile networks are characterized by highly critical and heterogeneous use cases that include \acf{eMBB}, \acf{URLLC}, and \acf{mMTC} \cite{song2020dynamic}. Among these use cases, the URLLC is the most challenging since it has stringent delay requirements to enable disruptive applications such as \ac{AR} and \ac{VR}, robot motion with closed-loop control for Industry 4.0, \ac{UAV} traffic control, remote surgery, and other critical time-sensitive applications \cite{el2015ngmn,alliance2019verticals,alliance20195g}. Envisioning, developing, and employing new technologies to enable such use cases are paramount for 5G and \ac{B5G} networks. One of these technologies is network slicing, a network concept described by \ac{NGMN} in \cite{el2015ngmn} and foreseen as one of the main drivers to permit a 5G network to reach all its potential \cite{afolabi2018network}. Network slicing enables the creation of multiple virtual networks that run simultaneously on the same physical network, allowing heterogeneous services to work concurrently on top of the operator's infrastructure with centralized resource orchestration \cite{yan2019intelligent,liu2021onslicing}. 

In order to feature end-to-end network slicing, the \ac{SMO} offers resource allocation for \acf{RAN}, \acf{TN}, and \ac{CN} \cite{O-RAN.WG1}. However, managing resources for heterogeneous network slices with different and strict requirements is challenging, mainly involving decision-making with high accuracy simultaneously for all domains --- \ac{RAN}, \ac{TN}, and \ac{CN} \cite{liu2021onslicing}. Regarding the RAN domain, different from legacy mobile networks, 5G leverages \ac{SDN} concepts to disaggregate RAN in \acf{RU}, \acf{DU}, and \acf{CU} nodes interconnected by multiple interfaces --- fronthaul, midhaul, and backhaul --- each one with its respective capacity and delay requirements \cite{foukas2016flexran,chang2017flexcran,garcia2018fluidran,garcia2018wizhaul}. It implies that the \ac{RAN} stack is divided into \acp{VNF} to be deployed at the \ac{RAN} nodes, which may be interconnected through different possible paths and by multiple nodes over the \ac{TN}. Complexity increases as \ac{SMO} deals jointly with \ac{RAN} slicing, flexible allocation of multiple \acp{VNF} to disaggregated \ac{RAN} nodes, and routing paths over the transport network while meeting \ac{QoS} requirements \cite{afolabi2018network}.

In the literature, \ac{RAN} slicing and flexible allocation of \acp{VNF} to \ac{RAN} nodes (i.e., \ac{VNF} placement) are commonly approached separately. The former has been the goal of several works \cite{foukas2017orion, yan2019intelligent, bega2019deepcog, tang2019service, li2020end, liu2021onslicing, murti2022constrained, adamuz2022stochastic}, as well as the latter \cite{foukas2016flexran, chang2017flexcran,garcia2018fluidran,garcia2018wizhaul, zhou2019utility}. It is also worth citing those works that jointly deal with \ac{VNF} placement and routing \cite{molner2019optimization, morais2022placeran, he2023leveraging}. Although many issues related to network slicing and \ac{VNF} placement have been identified and addressed separately, the joint approach is a more promising tool, as it takes into account more realistic network scenarios, enabling solutions to complex and real-world use cases. However, few works have presented contributions using this joint approach \cite{ojaghi2019sliced, matoussi2020user, ojaghi2022impact, motalleb2022resource}.
These articles are not tailored to assure end-to-end delay guarantees for \ac{URLLC} --- the most challenging slicing use case. On the one hand, the works in \cite{ojaghi2019sliced, matoussi2020user, ojaghi2022impact} deal with the delay in a simplified manner, i.e., without a formal queueing model to estimate end-to-end delay. On the other hand, the work in \cite{motalleb2022resource} uses a traditional queueing model based on average measures, such as average queue size and average waiting time \cite{little1961proof,kleinroch1976queueing,doshi1986queueing}, to estimate end-to-end delay. In this context, \ac{NC} emerges as a suitable alternative, a technique based on the worst-case approach \cite{cruz1991calculus,fidler2010survey}, providing the required queueing toolset to estimate backlog and delay bounds in full compliance with the \ac{URLLC} demands~\cite{she2017rrmurllc,bennis2018urllc}.

\textbf{Contributions.} In this work, we introduce a problem formulation that assures end-to-end delay for users of \ac{URLLC} slices by jointly choosing traffic routes, bandwidth allocation, and functional splits. 
To the best of our knowledge, our approach is the first to tackle resource allocation and routing for the transport network in a sliced and disaggregated RAN, providing delay guarantees to \ac{URLLC} users while assuring minimum throughput for \ac{eMBB} users. Our optimization model can accommodate multiple functional splits and different use cases with their respective requirements. The main contributions of this article are summarized in the following:

\begin{itemize}
    \item \textbf{A joint approach of network slicing, functional split, and routing} - for \ac{RAN} \ac{VNF} placement and resource allocation to the transport network that interconnects the three independent nodes --- \ac{RU}, \ac{DU}, and \ac{CU} --- of a disaggregated \ac{RAN}; 
    \item \textbf{A delay-constrained problem formulation} -  to the aforementioned joint approach to address complex real-world situations, such as multiple slices, multiple subnet instances per slice, and functional splits with different capacity and delay requirements in the same RAN unit;
    \item \textbf{A closed-form expression} - for the worst-case end-to-end delay of the \ac{RAN} transport network to be used in the aforementioned delay-constrained problem formulation;
    \item \textbf{A cash flow analysis} - to interrelate \ac{MNO} revenues and operating costs to be used in the objective function that maximizes \ac{MNO}'s profit;
    \item \textbf{A comprehensive performance evaluation} - using well-known, free, extensible, and component-based software tools: OMNeT++, DiscoDNC, and IBM CPLEX. 
\end{itemize}

\textbf{Article organization.} The remainder of this article is organized as follows: in Section 2, we present the background knowledge comprising network slicing, \ac{RAN} functional splitting, \ac{URLLC}, and network calculus basics; in Section 3, we present the system model, and we formulate the problem as a delay-constrained optimization problem with an objective function to maximize \ac{MNO}'s profit; in Section 4, first we present the performance evaluation method and the parameters setup, then we evaluate the performance of our proposal in comparison to baseline approaches; in Section 5, we review important works from the literature, and we highlight the most relevant findings and, finally, in Section 6, we present final considerations.
\section{Background}\label{sec:bg}

In the following, we present the theoretical foundations of our work, which are related to RAN slicing, RAN disaggregation, \ac{URLLC}, and network calculus.


\subsection{RAN Slicing}\label{sec:bg-slicing}
Network slicing is a fundamental technology for 5G that provides the ability to instantiate logical networks for different services with individual characteristics and requirements \cite{ojaghi2019sliced}. Network slices require logical isolation and intelligent resource allocation while running in the same physical infrastructure \cite{foukas2017orion,ojaghi2022impact}. This imposes orchestration issues when sharing limited resources among different slices while simultaneously meeting diverse application requirements. 
To offer guidelines to address these issues, the O-RAN Alliance defined an \ac{SMO} for 5G networks underlying the network slicing concepts present in \cite{3gpp.28.530}, where the \ac{NSMF} provides an end-to-end solution, allocating all required network resources during the lifecycle of an \ac{NSI}. NSI is defined as the set of one or multiple \acp{NSSI} providing all required resources, such as \acp{VNF}, transmission and processing capacities, for \ac{RAN}, \ac{TN}, and \ac{CN} domains, enabling end-to-end slice deployment over the physical infrastructure~\cite{O-RAN.WG1,3gpp20173gpp}. From the \ac{RAN} perspective, it means that both \ac{RAN} and TN nodes --- interconnecting the \ac{RAN} nodes --- are settled to satisfy all service requirements associated with the \ac{NSI} \cite{ordonez2021rollout}.

In general, \ac{TN} supports \ac{PDU} sessions over Ethernet \cite{O-RAN.WG4}. These \ac{PDU} sessions provide the paths to the data traffic between the \ac{5GC} and the \ac{UE}. From the network slicing perspective, \ac{QoS} assurance and traffic isolation depend on the operators correlating \ac{PDU} sessions to the unique identifiers of \acp{NSI} \cite{ferrus2018management}. It means that for each network operator, uniquely identified by the \ac{PLMN-ID}, \ac{PDU} session establishment between \ac{5GC} and UE depends on a network slice identification accomplished using the \ac{S-NSSAI} \cite{O-RAN.WG1}. Thus, each operator (identified by a \ac{PLMN-ID}) supports multiple slices, each one identified by a \ac{SST} and a \ac{SD} \cite{O-RAN.WG1}. For example, consider two \ac{URLLC} instances running in the same network operator's infrastructure but with distinct delay requirements. Although both instances share the same \ac{PLMN-ID}, they have distinct \ac{S-NSSAI} associated to \ac{PDU} sessions. 

In summary, in the 5G network slicing concept, data traffic between \ac{UE} and \ac{5GC} is possible using a slice instantiation uniquely identified by a \ac{S-NSSAI}. This instance has the required VNFs and network resources to allow the establishment of one or more \ac{PDU} sessions between \ac{UE} and \ac{5GC}. Moreover, the slicing concept is not limited to offering multiple services on the same operator's infrastructure. Multiple slices are also deployed for the purpose of obtaining RAN sharing between different network operators \cite{O-RAN.WG1}. \ac{RAN} sharing leverages the \ac{RAN} disaggregation to enable RAN nodes and resources --- including those provided by different vendors --- to be shared between multiple operators. Such flexible deployments are fully dependent on the RAN architecture.

\subsection{RAN Architecture}\label{sec:bg-split}




In the \ac{3GPP} 5G architecture, the \ac{NG-RAN} covers either a \ac{ng-eNB} or a \ac{gNB} connecting \ac{UE} to the \ac{5GC} through an NG interface. The ng-eNB provides legacy \ac{LTE}/\ac{E-UTRA} user and control planes, while gNB provides 5G \ac{NR} user and control planes \cite{3gpp.38.300}. The \ac{NG-RAN} leverages the concept of RAN disaggregation, in which RAN nodes --- RU, DU, and CU --- are interconnected through a transport network with multiple forwarding nodes demanding routing decisions. \acp{VNF} are placed on RAN nodes according to one or two-tier functional splits \cite{larsen2019,gstr2018transport}. RAN disaggregation with a \acf{FFS}, i.e. RAN deployment with functional splits (one or two-tier) chosen from multiple available functional split options, allows different levels of \textit{cloudification} and centralization, enabling the deployment of diverse scenarios with heterogeneous requirements \cite{habibi2021towards}. 

\begin{figure}[htb]
    \centering
    \includegraphics[width=0.45\textwidth]{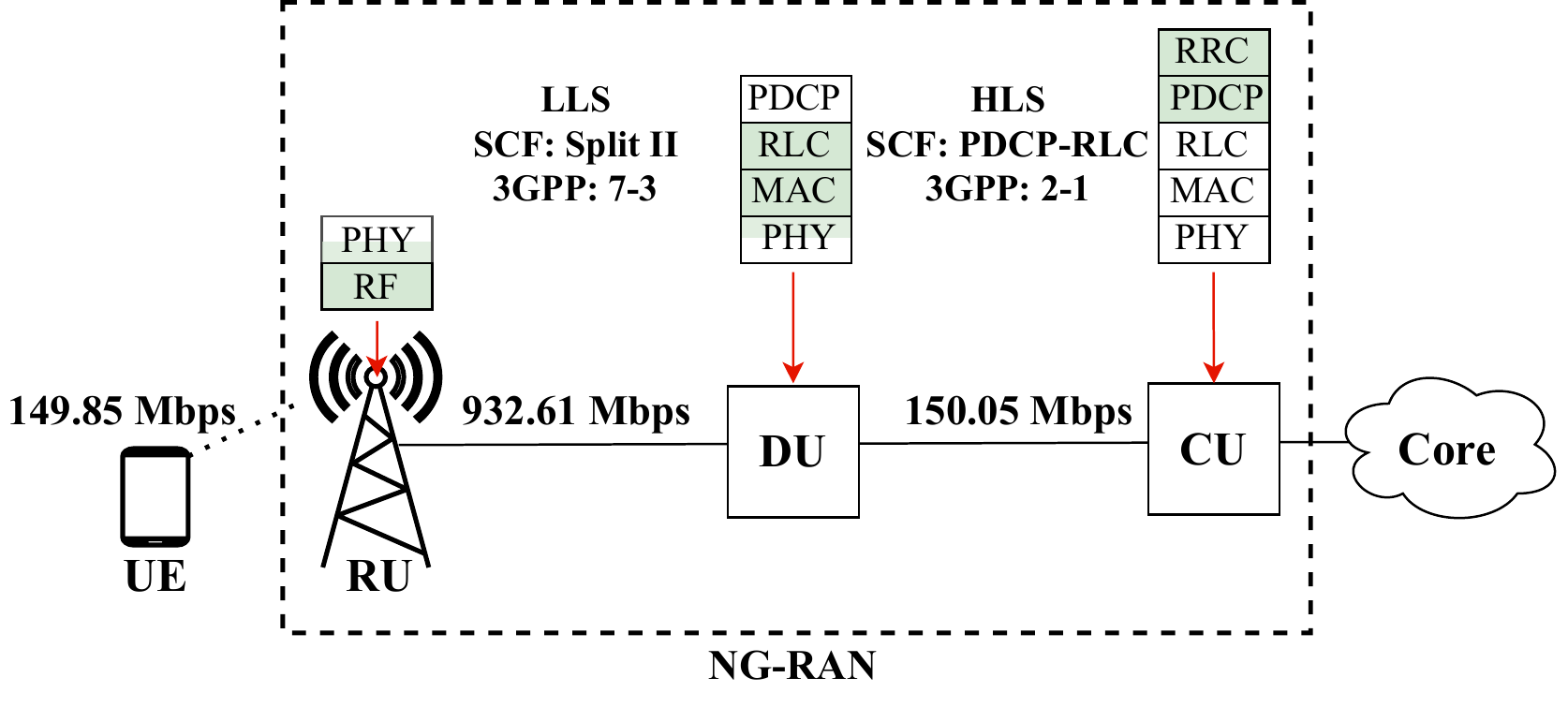}
    \caption{Disaggregated RAN example with LLS and HLS.}
    \label{fig:NG-RAN}
\end{figure}

\begin{table}[htbp]
\addtolength{\tabcolsep}{-2pt}
    \centering    
    \caption{Functional Splits for different organizations - Downlink}
        \label{tab:split}

    \begin{threeparttable}

    \begin{tabular}{|c|l|c|c|c|}
    \hline
         \textbf{Split} & \textbf{3GPP} \cite{3gpp.38.801} & \textbf{SCF} \cite{SCF.159.07.02} & \textbf{eCPRI} \cite{eCPRIV2.0} & \textbf{3GPP, SCF and eCPRI}\\
    \hline
    \hline\rowcolor{Gray}

        $O_1$  &   Opt. 1     &  RRC-PDCP           & A & $\checkmark$\\ \hline
        $O_2$  &  Opt. 2-1\tnote{*}    &  PDCP-RLC     & B& $\checkmark$\\ \hline\rowcolor{Gray}
        $O_3$  &   Opt. 3-1\tnote{**}     &  N/A          & N/A& \\ \hline        
        $O_4$  &   Opt. 4\tnote{***}     &  RLC-MAC      & C& $\checkmark$\\ \hline\rowcolor{Gray}
        $O_5$  &   Opt. 5     &  Split MAC    & N/A&  \\ \hline
        $O_6$  &   Opt. 6     &  MAC-PHY    & D& $\checkmark$\\ \hline\rowcolor{Gray}
        $O_7$  &   N/A     &     I   & N/A& \\ \hline
        $O_8$  &   Opt. 7-3   &  II          & I$_D$& $\checkmark$\\ \hline\rowcolor{Gray}
        $O_9$  &   Opt. 7-2   &  N/A         & II$_D$& \\ \hline
        $O_{10}$  &   Opt. 7-1   &  III         & N/A& \\ \hline\rowcolor{Gray}
        $O_{11}$  &   Opt. 8    &  IIIb           & E& $\checkmark$\\ \hline
        $O_{12}$  &   N/A    &  IV           & N/A& \\ 
        \hline
    \end{tabular}
      \begin{tablenotes}
   \item[*] In option 2-1, RRC and PDCP are both centralized for U-Plane and C-Plane. In option 2-2, PDCP is only centralized for U-Plane. 
   \item[**] In option 3-1, ARQ is implemented in DUs, while in option 3-2, it is centralized.
   \item[***] 3GPP did not see any benefit on using split option 4 \cite{3gpp.38.801}.
  \end{tablenotes}
      \end{threeparttable}
\end{table}
\begin{table*}[htb]
\addtolength{\tabcolsep}{-2.5pt}
    \centering    
    \caption{Capacity and delay requirements in the downlink direction for different functional split options (adapted from \cite{SCF.159.07.02})}
        \label{tab:splitchoiceSCF}
    \begin{threeparttable}

    \begin{tabular}{|l|l|c|}
    \hline
         \textbf{Split Option $O_j$}& \textbf{Capacity $\mu^{O_j}$ in Mbps} & \textbf{One-Way Delay $d^{O_j}$} \\
    \hline
    \hline \rowcolor{Gray}
        $O_1$ (3GPP option 1)  &   $IP_{DL}^{TTI}(IP_{pkt})N_{DL}^{TBS}\times 8\times 10^{-3}$    &  10 ms           \\
        \hline
        $O_2$ (3GPP option 2-1)  & $IP_{DL}^{TTI}(IP_{pkt}+Hdr_{PDCP})N_{DL}^{TBS}\times 8\times 10^{-3}$     &  1.5-10 ms    \\
            \hline\rowcolor{Gray}
        $O_4$ (3GPP option 4)  & $IP_{DL}^{TTI}(IP_{pkt}+Hdr_{PDCP}+Hdr_{RLC})N_{DL}^{TBS}\times 8\times 10^{-3}$     & 1 ms          \\   
            \hline
        $O_6$ (3GPP option 6)  &   $IP_{DL}^{TTI}(IP_{pkt}+Hdr_{PDCP}+Hdr_{RLC}+Hdr_{MAC})N_{DL}^{TBS}\times 8\times 10^{-3} +FAPI_{DL}$     &  250 $\mu$s / 2 ms\tnote{*}     \\
            \hline\rowcolor{Gray}
        $O_8$ (3GPP option 7.3) &   $(N_{UE}PDSCH_{REs}+PDCCH_{REs})N_{IQ}N_{L}\times10^{-3}$     &  250 $\mu$s / 2 ms\tnote{*}   \\
            \hline
        $O_{9}$ (3GPP(O-RAN) option 7.2(x))  &   $N_{SC}^{RB}\times N_{RB}\times N_{SYM}^{SUB}\times N_{L}\times N_{IQ}$     &  250 $\mu$s / 2 ms\tnote{*}  \\
            \hline\rowcolor{Gray}
        $O_{11}$ (3GPP option 8) &   $SR \times N_{AP} \times N_{IQ} $   &  250 $\mu$s / 2 ms\tnote{*}         \\
        \hline
    \end{tabular}
    \begin{tablenotes}
    \item [*] Near-ideal one-way latency assuming HARQ interleaving.
    \item $IP_{DL}^{TTI}=\frac{TBS_{DL}}{(IP_{pkt}+Hdr_{PDCP}+Hdr_{RLC}+Hdr_{MAC})\times 8}$.
    \item $PDSCH_{REs}=N_{RB}N_{SC}^{RB}(N_{SYM}^{SUB}-RefSym_{REs}N_{AP})$.
    \end{tablenotes}

      \end{threeparttable}
\end{table*}

Figure \ref{fig:NG-RAN} depicts an NG-RAN two-tier deployment, where a \ac{LLS} is deployed in the fronthaul --- interface connecting RU to DU, while a \ac{HLS} is deployed in the midhaul --- interface connecting DU to CU.
Generally, RUs are mainly responsible for \ac{RF} functions, and all other RAN functions are flexibly deployed at the DU or CU, imposing different delay and capacity requirements on the interfaces connecting them to the transport network \cite{larsen2019}. Figure \ref{fig:NG-RAN} illustrates the protocol stack for each RAN node, where those VNFs deployed at the respective RAN node are highlighted in green and those that are possible deployments when using other functional splits are in blank. We considered as NG-RAN VNFs the network functions both available in \ac{ng-eNB} and \ac{gNB}, i.e., \ac{PHY}, \ac{MAC}, \ac{RLC}, \ac{PDCP} and \ac{RRC}. Moreover, RAN network functions can be deployed and operated as VNFs on virtualized RAN nodes, i.e., \acp{vDU} or \acp{vCU}, leveraging the SDN paradigm with the RAN disaggregation concept \cite{murti2022constrained}. 
Additionally, we can observe the stringent bandwidth requirement for LLS in Fig. \ref{fig:NG-RAN}. In this context, LLS demands around six times more bandwidth than the UE requires, while HLS imposes a small overhead on the packets transmitted over the midhaul interface.

Concerning the functional split options, \ac{3GPP} considered a set of split options for the disaggregated NG-RAN architecture, as presented in Table \ref{tab:split} \cite{3gpp.38.801}. Similar studies were carried out by other standardization bodies, such as \ac{SCF} \cite{SCF.159.07.02} and \ac{eCPRI} \cite{eCPRIV2.0}, which are also presented in Table \ref{tab:split}, where $\mathcal{O}=\{O_1,O_2,\ldots,O_{12}\}$ is the set comprehending all split options. The functional split choice for each interface determines the requirements for delay and capacity. In this work, we use a subset of functional split options. For such a purpose, let $\mathcal{\hat{O}} \subset \mathcal{O}$ be the subset of functional split options considered simultaneously relevant by all three standardization bodies enlisted in Table \ref{tab:split} (3GPP, SCF, and eCPRI), i.e., 3GPP options 1, 2-1, 4, 6, 7-3, 8. Additionally, we add to $\mathcal{\hat{O}}$ the O-RAN LLS option 7.2x \cite{polese2022understanding} due to its paramount relevance to the \textit{de facto} industry standardization process, as presented in Table \ref{tab:split}. Therefore, $\mathcal{\hat{O}}=\{O_1,O_2,O_4,O_6,O_8,O_9,O_{11}\}$. The required capacity for each split option is calculated using the related expression enlisted in Table \ref{tab:splitchoiceSCF}. Moreover, we present in Table \ref{tab:splitchoiceSCF} the delay requirement for each split option.

On the one hand, we fixed split option 7.2x as the functional split for the LLS, in compliance with the O-RAN alliance. On the other hand, we allow a \ac{FFS} choice in HLS. In other words, while RAN LLS deploys O-RAN split option 7.2x, RAN HLS can be implemented with one of the following splits: 3GPP options 1, 2-1, 4, 6, or 7-3. By allowing this flexibility, we enable the deployment of a set of split options with different requirements for the midhaul interface, adding a new degree of freedom to the design and deployment of TN over the midhaul interface. In this work, we aim to tackle the RAN slicing in the context of a \ac{FFS}, considering that at the center of this problem lies one of the main issues for effective support of URLLC services in real-world deployments with disaggregated RANs.

\subsection{Joint RAN Slicing and Flexible Functional Split Approach}\label{sec:bg-joint}

In one approach, network slicing is a technology that can provide efficient resource reservation and logical network separation with features to meet the most demanding applications. In another approach, intelligent VNF placement at different RAN nodes can provide the flexibility demanded by a dynamic operation oriented to meet QoS requirements. Moreover, the dynamic queueing behavior at each network node and the adopted routing paths are relevant real-world concerns. Putting it all together is challenging, as we examine in the following.

\begin{figure*}\includegraphics[width=\textwidth]{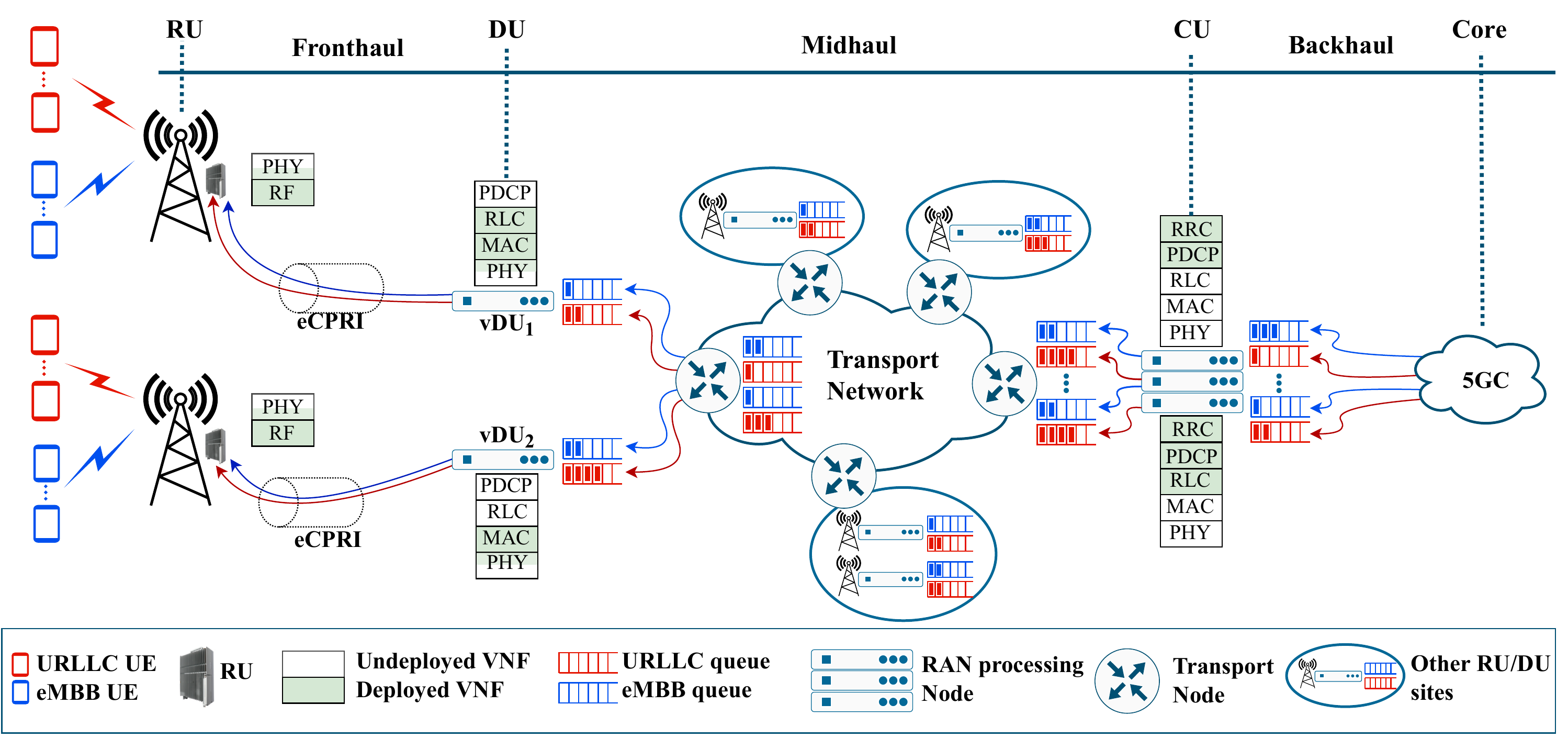}
	\caption{5G network scenario with network slicing, functional split, routing, and queueing system.}
	\label{fig:queueingModel}
\end{figure*}

Figure \ref{fig:queueingModel} illustrates a network scenario with a joint approach to network slicing, functional splitting, routing, and queue modeling. Multiple UEs are connected to each RU and demand traffic flows associated with different services provided by different NSIs (URLLC and eMBB). Multiple RUs are connected to the DU through a fronthaul interface, where data frames are transported using eCPRI. DU resources are virtualized in vDUs, and each vDU is associated with a single RU. A vDU serves multiple queues associated with NSIs. Different flows belonging to the same NSI enter the same queue at a vDU. However, flows associated with the same NSI, but related to other RUs, and consequently different vDUs, can be served with distinct functional splits. At the vDU, queues are served with different priorities --- shares of the transmission capacity --- based on the NSI requirements. 

Each set of flows belonging to the same NSI is forwarded from the \ac{5GC} to the \ac{CU} through the backhaul interface. Such network flows undergoing a specific functional split can be routed with different priorities per transport node throughout the midhaul interface --- where the traffic of other RU/DU sites are also routed --- from CU to vDUs, where the other VNFs of the chosen functional splits are deployed. In summary, Fig. \ref{fig:queueingModel} depicts the following issues:

\textbf{NSI in a disaggregated RAN.} Regarding the RAN domain, different from the full-stack deployment in legacy networks, RAN nodes can be localized in different sites interconnected by a transport network, implying that for a single NSI, the RAN domain may need multiple NSSIs with specific demands for VNF deployment, routing, and bandwidth allocation. When we consider scenarios with multiple NSIs with different requirements competing for the same resources --- which is a typical scenario envisioned by O-RAN in multi-operator RAN sharing in 5G \cite{O-RAN.WG1} --- we have a problem with a non-trivial solution.

\textbf{Multiple vDUs per NSI.} A vDU implements only a single functional split for its associated RU. Therefore, meeting the service requirements demanded by UEs can be challenging since traffic flows from the same NSI may be processed in different vDUs with different functional splits. Observe in Fig. \ref{fig:queueingModel} that the UEs in red use a service classified in the URLLC use case. However, considering the URLLC slice flows, some are processed in vDU$_1$ with a different functional split from that implemented by vDU$_2$, which processes another set of URLLC slice flows. All of them belong to the same NSI. 

\textbf{Multiple NSIs per vDU.} vDU$_1$ and vDU$_2$ process traffic flows from different NSIs (URLLC and eMBB). As other functional splits have mismatched latency and throughput requirements, meeting application requirements while considering a sliced and disaggregated RAN deployment is a challenging task. This becomes even more demanding when a URLLC instance is involved because the URLLC use case depends on meeting strict end-to-end latency requirements. 

\textbf{Queueing delay in an NSI.} In addition to processing, transmitting, and propagation delays, real-world deployments must be aware of queueing delays since in a disaggregated RAN --- due to the multiple nodes --- the sum of all queueing delays can critically impact the overall performance. Time-sensitive applications, e.g. URLLC use cases, have strict delay requirements. Such concern is even more critical when URLLC traffic competes for resources with other slices. For a \acf{FFS}, overheads are added to the URLLC packets when VNFs are placed in the CU \cite{larsen2019}, affecting the total delay budget. Dealing with this issue is not straightforward because URLLC application requirements are not fulfilled by average performance but by guarantees of maximum delay.

\subsection{URLLC}\label{sec:bg-urllc}

URLLC is an enabler use case for the majority of high-expected applications and deployment scenarios in 5G networks, such as \ac{V2X}, \ac{IoT}, Industry 4.0, robot remote control, Augmented and Virtual Reality (AR/VR), tactile applications, and others \cite{kim2018ultrareliable}. Such applications are generally associated with automatic control with real-time interactions, and sensory-related use cases \cite{alliance2019verticals}, deserving special attention since they have, besides typical \ac{SLA} specifications, stringent delay and reliability requirements. The fulfillment of all URLLC demands, in full compliance with 3GPP standards, is one of the most challenging targets for 5G and B5G networks (actually, even stricter requirements are expected for 6G systems), playing an essential role in network design, deployment, and management.

Regarding the network traffic profile, in contrast with eMBB and mMTC use cases, many URLLC services can present smaller packet payloads, and the packet arrival is usually assumed to follow a constant inter-arrival time with a fixed packet size \cite{gao2021study}. 
While a small packet presents a minor processing delay, the overhead to transmit these small packets over the transport network connecting disaggregated RAN nodes can compromise the delay budget \cite{alliance2019verticals}. For example, in a typical industry automation case, a central unit transmits real-time control data, composed of small packets, to control a robot with a time-critical function, requiring information to be delivered to the robot with low latency and high reliability \cite{kalor2018network}. In some cases, tactile information will be fed back to guide the upcoming control commands. Such forward and feedback information requires guarantees, establishing some challenging issues for URLLC deployment in critical applications and opening up perspectives on how 5G and B5G systems will be able to address all URLLC strict requirements on real-world applications \cite{alliance20195g}. The upcoming solution is not trivial and involves physical layer enhancements, a well-planned transport network, end-to-end architecture flexibility, and trade-offs in business models \cite{alliance2019verticals}.

A disaggregated and sliced RAN poses additional difficulties in supporting the URLLC use case. For example, although URLLC and eMBB have different characteristics and demands, they may need to coexist in the same RAN deployment, competing simultaneously for resources of RAN processing and transport network nodes. Indeed, when URLLC use case comes into play, every factor impacts end-to-end delay and reliability guarantees. Naive solutions based on throughput increase or similar one-size-fits-all solutions cannot handle the heterogeneity of this problem, comprehending multiple split options, network topologies, user traffic demands, decisions between edge or centralized computing, and related operational costs \cite{alliance20195g}. In this scenario, the adopted queueing model is an important aspect to keep in mind to guarantee no violations of the delay requirement. Classical queueing theory is usually applied, assuming probabilistic models for the network traffic. The stringent URLLC requirements do not accept such modeling uncertainties \cite{kalor2018network}, leading us to alternative approaches based on worst-case analyses, such as that provided by \ac{DNC} \cite{le2001network}.

\subsection{Network Calculus}\label{sec:bg-netcalc}

Network calculus is a framework from queueing theory based on the min-plus algebra, providing closed-form expressions for backlog and delay bounds for data communications systems. These bounds are computed using functions for arrival and departure processes. Analogously to the linear systems theory, the relationship between the arrival and departure processes defines a function that describes the system behavior. In network calculus, this function is named a service curve, similar to an impulse response in a linear system \cite{fidler2010survey}.

\begin{table}[htb]
\centering
	\caption{Key Notations}
 \label{tab:keynotations}

	\begin{threeparttable}

	\begin{tabular}{|l|p{6cm}|}
            \hline
		\textbf{Symbol} & \textbf{Description}\\
            \hline
            \hline \rowcolor{Gray}
		$A,D$ & Arrival and departure processes, respectively. \\
		\hline
		$\alpha,\beta$&  Arrival and service curves, respectively.\\
		\hline\rowcolor{Gray}
		$d, q$&  Delay and backlog, respectively.\\
  \hline
  $d_{max}, q_{max}$& Worst-case bound for the delay and backlog, respectively.\\
	\hline\rowcolor{Gray}
$\mathcal{O},\mathcal{\hat{O}}$&  Set of all and eligible split options, respectively.\\
        \hline
		$\mathcal{B},\mathcal{F},\mathcal{S},\mathcal{U},\mathcal{G}$&  Set of RUs, flows, slices, vDUs and VNFs, respectively.\\
		\hline\rowcolor{Gray}
            $B,F,S,U,G$ & Number of RUs, flows, slices, vDUs and VNFs, respectively. \\
		\hline
		$R,T$&  Rate-latency parameters: transmission rate and fixed-latency, respectively.\\
		\hline\rowcolor{Gray}
		$\rho,\sigma$&  Arrival curve parameters: arrival rate and burst, respectively.\\
		\hline
		$\Pi, \mathscr{R},\mathscr{C}$&  Profit, revenue and cost, respectively.\\
  \hline\rowcolor{Gray}
  	$\Psi_{\mathscr{R}},\Psi_\mathscr{C}$& Total revenue and total cost, respectively.\\
  \hline
		$\mathscr{C}_g^{CU}, \mathscr{C}_g^{DU}$& Cost to deploy and run VNF at the CU and vDU, respectively.\\
  \hline\rowcolor{Gray}
		$K_0, K^u$& Number of RCs at the CU and vDU, respectively.\\
		\hline
            $U^s$&  Number of vDUs related to the $s$-th slice.\\
		\hline \rowcolor{Gray}
		$F^s$& Number of flows in the $s$-th slice.\\
		\hline
		$F^{s,u}$&  Number of flows in the $s$-th slice instance related to the $u$-th vDU\\
		\hline \rowcolor{Gray}
   $F_{BE},F_{max}$ &  Number of UEs generating revenue in the network at the \ac{BE} point and at full capacity, respectively. \\
 \hline
		$f_i^{s,u}$&  $i$-th flow of the $s$-th slice processed in the $u$-th vDU.\\
		\hline\rowcolor{Gray}
		$R_v,\beta_v,S_v$ &  Transmission rate, service curve and number of slices related to the $v$-th network node, respectively.\\
		\hline
		$w_v^{s,u}, \phi_v^{s,u}$& Assigned weight and percentage of transmission rate $R_v$ allocated to the $s$-th slice instance related to the $u$-th vDU, respectively.\\
	\hline\rowcolor{Gray}
	$d_{i}^{s,u}$,$d_{i,TN}^{s,u}$ & End-to-end and transport delay of the $i$-th UE associated with the $s$-th slice and $u$-th vDU.\\
 \hline
	$d_{i,RAN}^{s,u}$,$d_{i,PD}^{s,u}$ & RAN and PD delay of the $i$-th UE associated with the $s$-th slice and $u$-th vDU.\\
  \hline\rowcolor{Gray}
  $d_{SLA},\mu_{SLA}$ &  QoS requirements: maximum delay and minimum throughput, respectively. \\
     \hline
$a(d_{SLA},\mu_{SLA})$ &  Service $a$ with requirements: maximum delay $d_{SLA}$ and minimum throughput $\mu_{SLA}$.\\
 \hline\rowcolor{Gray}
 $FoI$ &  Flow of Interest, i.e., the flow under analysis.\\
 \hline
 $\beta^{l.o.f}$ &  Left-over service curve for flow $f$. \\
 \hline\rowcolor{Gray}
 $SR$ &  Sample Rate. \\
 \hline
 $N_L$,$N_{AP}$ &  Number of layers and antenna ports, respectively. \\
 \hline\rowcolor{Gray}
 $N_{SC}^{RB}$, $N_{SYM}^{SUB}$ &  Number of subcarriers per RB, and symbols per subcarrier, respectively. \\
 \hline
 $N_{RB}$, $N_{IQ}$ &  Number of RBs and  bitwidth, respectively. \\
 \hline\rowcolor{Gray}
  $\eta$ &  Ratio between VNF deployment cost at the CU and vDU. \\
 \hline
 $\gamma$ &  Ratio between revenue per UE and cost per VNF deployment at a vDU. \\
 \hline\rowcolor{Gray}
 $\zeta$ &  \acf{BE} point. \\
 \hline
 $\mathcal{P}^{s,u}$,$\mathcal{P}^{s,u}_{l}$ & Set of all available paths and set of transport nodes in the chosen routing path to the $s$-th NSI and $u$-th vDU, respectively.\\
 \hline\rowcolor{Gray}
 $V^{s,u}_{l}$ & Number of transport nodes in the chosen routing path of the $s$-th NSI and $u$-th vDU.\\
 \hline
\end{tabular}
 \end{threeparttable}
\end{table}

\begin{figure}[hbt!]
    \centering
\includegraphics[width=6cm]{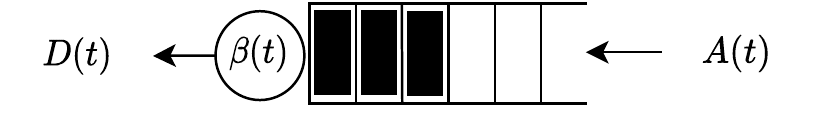}    \caption{Queue-server system with arrival process $A(t)$, service curve $\beta(t)$ and departure process $D(t)$.}
    \label{fig:queue}
\end{figure}

Figure \ref{fig:queue} illustrates a queue-server system where $A(t)$, $\beta(t)$, and $D(t)$ are, respectively, the arrival process, service curve, and departure process. Key notations used in this work are summarized in Table \ref{tab:keynotations}. Formally, the following definitions are building blocks for all network calculus formulations.

\begin{definition}[Arrival Curve]
\label{def:arrival_curve}
The arrival curve $\alpha(t)$ is an upper bound for the arrival process $A(t)$, as follows \cite{le2001network}:
\begin{equation}
    A(\tau,t)\leq \alpha(t-\tau),  t\geq \tau \geq 0, \forall t.
    \label{eq:def1}
\end{equation}
\end{definition}
\begin{definition}[Service Curve]
\label{def:service_curve}
A system with an arrival process $A(t)$ and departure process $D(t)$ has service curve $\beta(t)$ for which the following inequality holds for all $t\geq 0$:
\begin{equation}
    D(t)\geq \inf_{0\leq \tau \leq t}\{A(\tau)+\beta (t-\tau)\}=A\otimes \beta (t).
    \label{eq:def2}
\end{equation}
\end{definition}

\begin{figure}[htb!]
    \centering
\includegraphics[width=8.5cm]{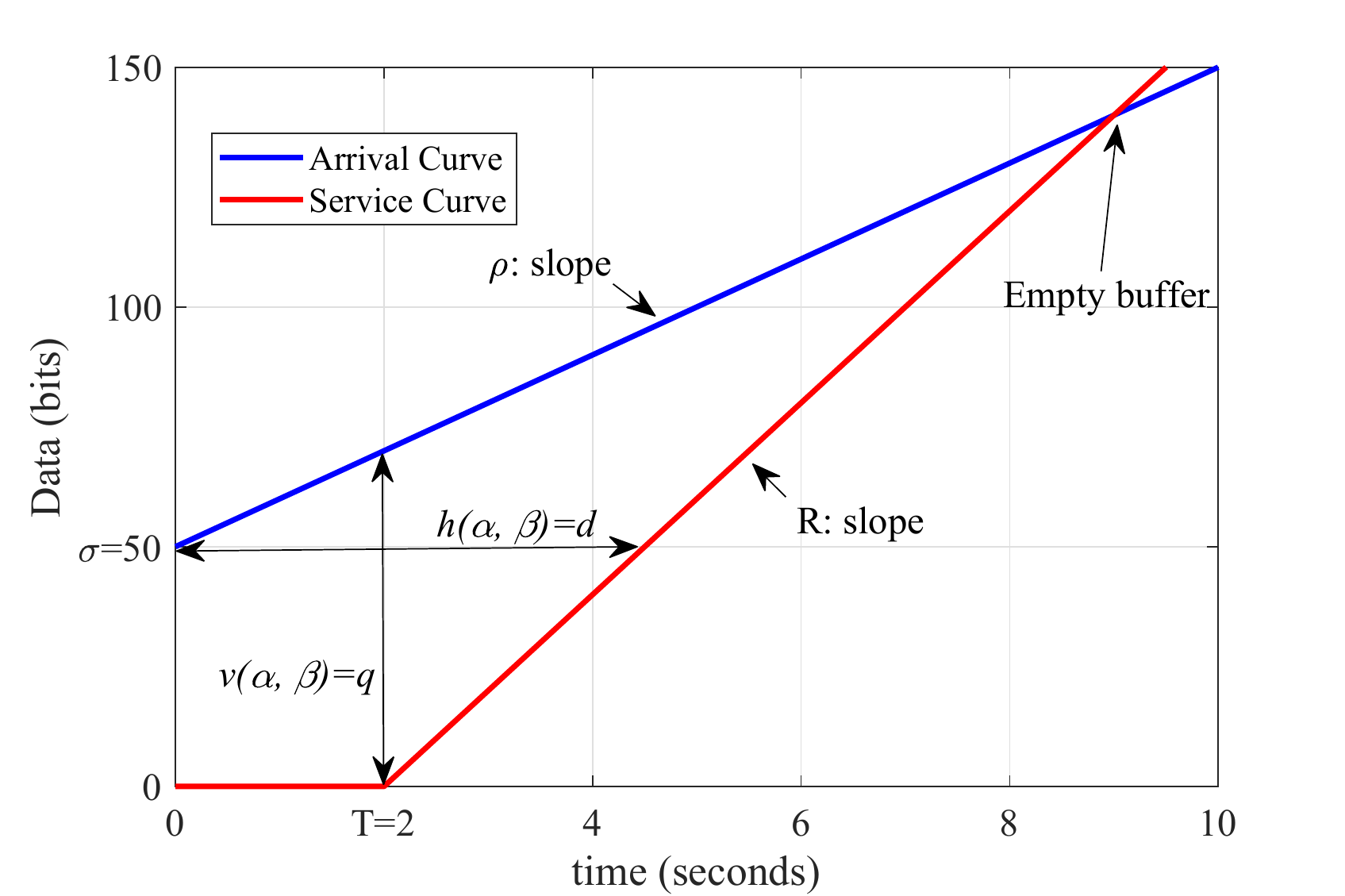}    
\caption{Estimating backlog and delay based on the arrival and service curves.}
    \label{fig:ncb}
\end{figure}


Figure \ref{fig:ncb} depicts a theoretical example where the arrival curve $\alpha(t)$ (also known as the envelope process) is a leaky-bucket function, and the service curve $\beta(t)$ is a rate-latency function, as follows:
\begin{equation}
    \alpha(t)=\rho t+\sigma,
    \label{leaky-bucket}
\end{equation}
\begin{equation}
    \beta(t)=R [t-T]^+.
    \label{rate-latency}
\end{equation}
\noindent where (\ref{leaky-bucket}) is a monotonic non-decreasing function or monotonic wide-sense increasing function, $\rho$ is the data rate, $\sigma$ data burst. In (\ref{rate-latency}), $R$ is the service rate, and $T$ is a fixed latency which can be used to denote, for example, the delays related to packet serialization, switching fabric, and packet processing \cite{mathew2020network}. The operator $[x]^+$ represents $\max(x,0)$. For the Deterministic Network Calculus, (\ref{leaky-bucket}) is a strict bound for the input data traffic, i.e., for all considered time intervals, no violations to  (\ref{leaky-bucket}) are observed. However, (\ref{rate-latency}) denotes the minimum rate $R$ offered to the input flow with maximum fixed latency $T$. In Fig. \ref{fig:ncb}, the vertical distance  ($v(\alpha,\beta)$) between curves $\alpha(t)$ and $\beta(t)$ represents the backlog (queue length in the buffer), while the horizontal distance ($h(\alpha,\beta)$) represents the delay. The largest vertical and horizontal distances represent the worst-case backlog and delay, respectively. 

By assuming that a leaky-bucket function can describe the arrival curve, we are also considering that data can be generated in bursts but enter the network at regular intervals. Therefore, we assume a buffer size large enough where losses are not recorded. For every instant $t$, when $A(t) > D(t)$, there will be a non-empty queue in the buffer, and, for system stability, $R\geq \rho$. The system's server transmits data, whenever available, at a constant rate $R$ (work-conserving assumption). In this scenario, a fluid system is assumed in which the data is divisible down to the basic unit (bit) of the transmitted data. 

Based on the definitions mentioned earlier, closed-form equations for backlog and delay bounds are stated in the following theorems. For conciseness, we do not show proofs here, but they can be found in \cite{le2001network}.

\begin{theorem}[Backlog Bound]
\label{theo:backlog_bound}
The backlog of a system with arrival process $A(t)$ and departure process $D(t)$ is calculated as follows \cite{le2001network}:  
\begin{equation}
	q(t)=A(t)-D(t),
 \label{eq:backlog}
\end{equation}

\noindent
taking into account definitions \ref{def:arrival_curve} and \ref{def:service_curve}, one may insert (\ref{eq:def1}) and (\ref{eq:def2}) in (\ref{eq:backlog}) to obtain the worst-case bound for the backlog as follows:
\begin{equation}
    q_{max}\leq \sup_{0 \leq \tau \leq t}\{\alpha(\tau)-\beta(\tau)\}.
    \label{eq:q_max}
\end{equation}
\end{theorem}

\begin{theorem}[Delay Bound]
\label{theo:delay_bound}
Based on the min-plus algebra, the delay process $d(t)$ in a system with arrival process $A(t)$ and departure process $D(t)$ is estimated as follows \cite{le2001network}:
\begin{equation}
	d(t)=\inf_{0 \leq \tau \leq t}\{\tau \geq 0: A(t-\tau)\leq D(t)\},
 \label{eq:delay}
\end{equation}
\noindent taking into account definitions \ref{def:arrival_curve} and \ref{def:service_curve}, one may insert (\ref{eq:def1}) and (\ref{eq:def2}) in (\ref{eq:delay}) to obtain the worst-case bound for the delay:
\begin{equation}
	d_{max}\leq \sup_{t \geq 0}\{\inf\{\tau \geq 0: \alpha(t-\tau)\leq \beta (t)\}\}.
\end{equation}
\end{theorem}

Taking into account the previously mentioned definitions and theorems, and assuming \ac{FIFO} as queueing discipline, $d_{max}$ is calculated using Theorem \ref{theo:delay_bound} for a single flow in a single queue-server system as follows \cite{le2001network}:
\begin{equation}
\label{eq:delaymax}
    d_{max}=T+\frac{\sigma}{R}.
\end{equation}
Note that (\ref{eq:delaymax}) depends on the two parameters of the service curve ($R$ and $T$), but it only depends on the parameter $\sigma$ of the arrival curve, i.e., the $d_{max}$ estimation does not depend on $\rho$.

\textbf{Tandem systems.} 
The network calculus theory becomes particularly useful in systems with multiple nodes, where each node constitutes a queue-server system. For example, in Fig. \ref{fig:queueingModel}, from RU to CU, each node serves multiple slices, once the service curves for each slice in each of the $V$ nodes are known, it is possible to model the entire network using the service curve of the network $\beta_{net}(t)$, obtained through the network calculus property known as concatenation, which consists of the min-plus convolution of the service curves of $V$ nodes \cite{fidler2010survey}:
\begin{equation}
    \beta_{net}(t)=\beta_0(t)\otimes \beta_1(t)\otimes \dots \otimes \beta_{V-1}(t).
    \label{eq:conc}
\end{equation}
Once the network is modeled through the $\beta_{net}$ function, the knowledge of the arrival curve $\alpha(t)$ is sufficient to estimate the end-to-end delay of the network. This estimate is independent of the order of the servers in the tandem system, so identifying a bottleneck limiting network performance is crucial. However, its position in the network does not change the estimate \cite{fidler2010survey}.

For a tandem system where $V$ servers have the same service curve, that is, $\beta(t)=R[t-T]^+$, the network service curve is $ \beta_{net}(t)=R[t-VT]^+$ and the worst-case delay is obtained using the concatenation property:
\begin{equation}
    d_{max}= VT+\frac{\sigma}{R}.
\label{eq:concatenation}
\end{equation}
Using the additive method, the departure process of the $v$-th server will be the arrival process of $v+1$-th server. Therefore, the arrival curve of the $v$-th server is defined as:
\begin{equation}
\alpha(t)=\sigma + \rho t + v\rho T.
\end{equation}
The worst-case delay using the additive method is calculated as follows:
\begin{equation}
        d_{max}= VT+\frac{V(\sigma+\frac{V-1}{2}\rho T)}{R}.
\label{eq:additive}
\end{equation}
Burst is taken into account $V$ times in (\ref{eq:additive}) to calculate $d_{max}$ using the additive method, resulting in a very conservative estimate for the worst-case delay. Through the concatenation property, the burst is used only once to estimate the delay in (\ref{eq:concatenation}), coining the term \ac{PBOO}. In this case, PBOO is always valid for queue-server systems with FIFO discipline.

\textbf{Packetizer.}
When modeling a packet network as a fluid system, inaccuracies can be observed when estimating QoS parameters, e.g., when analyzing the influence of packet size on network delay. For example, consider our scenario depicted in Fig. \ref{fig:queueingModel}, where URLLC and eMBB compete for resources along the network. These two slices have, in general, different packet sizes. While small packet sizes characterize URLLC applications,  eMBB are frequently described in the literature with the \ac{MTU} \cite{ordonez2021rollout}.  In network calculus, an approach aiming to acquire a more accurate model combines a fluid system with a packetizer. The service curve $\beta_p(t)$ for such a system (tandem system composed of a fluid model and a packetizer) is defined as follows:
\begin{equation}
     \beta_p(t)=[\beta(t)-l_{max}]^+,
\end{equation}
\noindent where $\beta(t)$ is the fluid system service curve and $l_{max}$ is the size in bits of the maximum packet size --- a common approach for systems with non-preemptive servers with static priority. In the most pessimistic scenario, the highest priority flow will be served at the end of the transmission of the packet with the maximum size and lowest priority \cite{le2001network}.





\section{System Model and Problem Formulation}
\label{sec:SystemModel}

This section initially defines our system model, considering multiple \acp{UE}, \acp{RU}, \acp{NSI}, functional splits, and routing paths throughout the TN. Subsequently, we formulate an optimization problem that maximizes \ac{MNO}'s profit, detailing what is required to have a realistic relationship, in monetary units, between revenue and cost. Finally, based on \acf{DNC}, we derive a closed-form equation that estimates an end-to-end delay upper-bound to be used as a constraint in our optimization model.

\subsection{System Model}

The system model considers the \ac{NG-RAN} disaggregation into \ac{RU}, \ac{DU}, and \ac{CU}. Figure \ref{fig:queueingModel} illustrates communication in the  downlink direction, where the \acp{RU} communicate directly with the \acp{UE} through a wireless communication channel. A \ac{UE} consumes data from an application service associated with a traffic flow through a PDU session. \acp{RU} are physically close and interconnected to the corresponding DU through a fronthaul interface. A DU can run multiple vDUs, and each RU is associated with a single vDU. Each vDU can process flows from different \acp{NSI} and flows from diverse \acp{RU} can be related to the same \ac{NSI}. Additionally, we adopt the \acf{FFS} approach, in which the midhaul interface connecting CU to the \acp{vDU} can implement distinct functional splits. \acp{NSI} for certain critical services (e.g., time-sensitive services such as \ac{URLLC}) can be given higher priority for using TN resources than non-delay-sensitive slices such as \ac{eMBB}. 
In this context, the system model can be defined as follows.

Let $\mathcal{B}$ be the set of $B$ \acp{RU} in a 5G network so that $b \in \mathcal{B}=\{1,\cdots,B\}$. Let $\mathcal{U}$ be the set of $U$ vDUs so that $u \in \mathcal{U}=\{1,\cdots,U\}$. Let $\mathcal{G}$ be the set of VNFs that can be deployed at the CU so that $g \in \mathcal{G}=\{1,\cdots,G\}$.
VNFs that are not deployed at the CU are deployed at the \acp{DU}. The number of VNFs deployed at the $u$-th vDU is related to the functional split option $O_j \in \mathcal{O}$. Let $\mathcal{S}$ be the set of $S$ \acp{NSI}, so that $s \in \mathcal{S}$. In our model, each \ac{UE} is exclusively associated with a single traffic flow of a service provided by an NSI so that the number of traffic flows in the network is equal to the number of UEs. Let $\mathcal{F}$ be the set of $F$ flows from all RUs and $\mathcal{F}^{s}\subseteq \mathcal{F}$ the subset with $F^{s}$ flows associated with NSI $s$. Furthermore, let $\mathcal{F}^{s, u}$ be the set of $F^{s,u}$ flows so that $f_i^{s,u} \in \mathcal{F}^{s, u}=\{f_0^{s,u}, f_1^{s,u},\cdots,f_{F^{s,u}-1}^{s,u} \}$ is the $i$-th flow of the $s$-th NSI related to the $u$-th vDU. Let $\mathcal{V}$ be the set of all $V$ transport nodes in the network so that $v_v \in \mathcal{V}=\{v_0,v_1,\cdots,v_{V-1}\}$, where $R_v$ is the capacity in bps of the $v$-th transport node. In our model, the transmission rate $R_v$ of each transport node $v$ is represented by its service curve, being agnostic about the number of network interfaces, albeit ensuring that the sum of all transmission resources allocated to all interfaces in $v$ does not exceed its capacity $R_v$. Let $\mathcal{P}^{s,u}$ be the set of $L$ available paths (each path is composed of $V^{s,u}_l$ transport nodes in $\mathcal{V}$) to route traffic flows of the $s$-th NSI from the CU to the $u$-th vDU throughout the transport network, where $\mathcal{P}^{s,u}_{l} \subseteq \mathcal{P}^{s,u}$ is the set of $V^{s,u}_l$ transport nodes of the $l$-th routing path, where $l=\{1,\cdots,L\}$. Finally, we assume a feed-forward network, i.e. an acyclic network topology \cite{fidler2010survey},  where no loops are present.



\subsection{Problem Formulation}

Our objective is to maximize MNO's Profit $\Pi$ based on the simple difference between total revenue $\Psi_{\mathscr{R}}$ and total cost $\Psi_{\mathscr{C}}$. 


\textbf{Revenue ($\mathscr{R}$).} We consider a network scenario with $s\in \mathcal{S}$ where $s=embb$ is an NSI demanding $\mu^{embb}_{SLA}$ as minimum  bandwidth per RU. The MNO's revenue per UE ($\mathscr{R}$) is obtained from UEs using $s=urllc$, which is an NSI related to the URLLC service $a$ with SLA requirements denoted as $a(d^{urllc}_{SLA},\mu^{urllc}_{SLA})$, where $d^{urllc}_{SLA}$ stands for the maximum allowed end-to-end delay and $\mu^{urllc}_{SLA}$ stands for the minimum required throughput per UE. SMO instantiates the $s$-th NSI commissioning resources needed by RAN and TN nodes to meet such demands. SMO also decides the VNF placement, i.e., in which RAN node the $g$-th VNF from the RAN protocol stack will be deployed. Concerning the TN nodes, SMO determines the resource share $\phi_{v}^{s,u}$ (percentage) allocated to the $v$-th node to serve the $s$-th NSI related to the $u$-th vDU. We define $\phi_{v}^{s,u}$ in function of the weights $w_v^{s,u}$, i.e. the weighted allocation of resources of the $v$-th transport node to the set of flows of the $s$-th NSI related to the $u$-th vDU, as follows:
\begin{equation}
 \phi_{v}^{s,u}=\dfrac{w_v^{s,u}}{\sum_{i=1}^{S_v}\sum_{j=1}^{U^{s}_v}w_v^{i,j}}, \forall v\in\mathcal{V},\forall s\in\mathcal{S},\forall u\in\mathcal{U},
 \label{eq:weights}
\end{equation}
\noindent where $S_v$ is the number of NSIs routed through the $v$-th TN node, and $U^{s}_v$ is the number of vDUs related to the $s$-th NSI routed through the $v$-th TN node.
Therefore, the set of traffic flows belonging to the $s$-th NSI related to the $u$-th vDU will be assigned transmission rate $R_v^{s,u}$ to the $v$-th TN node, as follows:
\begin{equation}
    R_v^{s,u}=\phi_v^{s,u}R_v, 0\leq\phi^{s,u}_{v}\leq 1,\forall v\in\mathcal{V},\forall s\in\mathcal{S},\forall u\in\mathcal{U}.
    \label{eq:phi}
\end{equation}

\textbf{Cost ($\mathscr{C}$).} In our network scenario, the LLS is O-RAN 7.2x between RUs and vDUs. All vDUs are connected to a single CU through the TN. HLS is flexible, i.e., VNF placement at vDU or CU implies different costs, being cost-saving to instantiate VNFs at the CU \cite{suryaprakash2015heterogeneous}.
For the sake of simplicity, we incorporate in $\mathscr{C}$ all RAN operational costs using the number of RAN VNFs deployed at the vDU or CU. In this context,  $\mathcal{G}$ is the set of RAN VNFs available for HLS, and let $x_g^u\in \{0,1\}$ be the decision variable determining whether the $g$-th RAN VNF related to the $u$-th vDU will be placed at the CU or not. Moreover, we establish $0 < \eta \leq 1$ as the ratio between VNF deployment costs in CU ($\mathscr{C}^{CU}_g$) and vDU ($\mathscr{C}^{DU}_g$), as follows:
\begin{equation}
    \mathscr{C}^{CU}_g=\eta \mathscr{C}^{DU}_g, \forall g \in \mathcal{G}
    \label{eq:eta}
\end{equation}
\noindent
formally stating there is a cost saving when a RAN function is centralized (CU) compared to a distributed deployment (vDU). 

\textbf{Cash Flow Analysis.} We need a relationship in monetary units between Revenue $\mathscr{R}$ per UE and Cost $\mathscr{C}$ per VNF deployment to obtain coherent Profit $\Pi$ estimations through our analyses. Therefore, we present Theorem \ref{theo:CFA} to state this relationship.

\begin{theorem}
\label{theo:CFA}
Let $F_{max}$ be the maximum supported number of simultaneous UEs generating revenue in the network of an MNO and $F_{BE}$ the number of admitted UEs to the network that provides break-even between total Revenue $\Psi_\mathscr{R}$ and total Cost $\Psi_\mathscr{C}$. In such a network, there is a $\gamma$ value relating Revenue per UE ($\mathscr{R}$) and Cost per RAN VNF deployment on a vDU ($\mathscr{C}^{DU}_g$), as follows:
\begin{equation}
        \gamma=\frac{\mathscr{R}}{\mathscr{C}^{DU}_g}, \forall g \in \mathcal{G}.
\end{equation}

\end{theorem}

\renewcommand\qedsymbol{$\blacksquare$}

\begin{proof}
The cost to deploy VNFs at the vDU or CU varies with each split option, being cost-saving to deploy VNFs at the CU. Therefore, an upper bound for the total cost can be estimated by deploying all VNFs (but the RRC) at the vDUs, using the so-called \ac{D-RAN} approach for all vDUs in the network (i.e., $O_1$ in Table \ref{tab:splitchoiceSCF}), as follows:
\begin{equation}
    \Psi^{max}_\mathscr{C}=U[(G-1)\mathscr{C}^{DU}_g+\mathscr{C}^{CU}_g].
    \label{eq:revenue4}
\end{equation}
Substituting (\ref{eq:eta}) in (\ref{eq:revenue4}), we obtain:
\begin{equation}
    \Psi^{max}_\mathscr{C}=U\mathscr{C}^{DU}_g[(G-1)+\eta].
        \label{eq:revenue4a}
\end{equation}
In this approach, $F_{BE}$ is the number of UEs connected to the network generating revenue for the MNO that provides a break-even (BE) point between total revenue and total cost. At this \ac{BE} point, we have: 
\begin{equation}
    \Psi_\mathscr{R}=\Psi_{\mathscr{C}}^{max}.
    \label{eq:revenue5}
\end{equation}
Hence, we state that there is a $\zeta \in (0,1]$ at the BE point, defined as:
\begin{equation}
    F_{BE} =\zeta F_{max}.
    \label{eq:revenue1}
\end{equation}
\noindent
At this point, it is straightforward that 
\begin{equation}
    \Psi_{\mathscr{R}}=\mathscr{R}F_{BE}.
    \label{eq:revenue2}
\end{equation}
Substituting (\ref{eq:revenue1}) into (\ref{eq:revenue2}), we obtain the following:
\begin{equation}
     \Psi_{\mathscr{R}}=\mathscr{R}\zeta F_{max}.
    \label{eq:revenue3}
\end{equation}
\noindent
We obtain $\gamma$, the ratio between revenue per UE and cost per RAN VNF deployment at a vDU, substituting (\ref{eq:revenue4a}) and (\ref{eq:revenue3}) in (\ref{eq:revenue5}), as follows:
\begin{equation}
    \gamma=\frac{\mathscr{R}}{\mathscr{C}^{DU}_g}=\frac{U(G+\eta-1)}{\zeta F_{max}}.
    \label{eq:gamma}
\end{equation}
\end{proof}


In this approach, $\mathscr{R}$ can be interpreted as a simplified form of \ac{ARPU}, a standard measure in the telecommunication industry, i.e., total revenue $\Psi_{\mathscr{R}}$ divided by the number of UEs. We must estimate $\zeta$ at the \ac{BE} point to obtain $\gamma$ using (\ref{eq:gamma}). For evaluation, we propose to estimate $\zeta$ based on the cash flow from the operating activities of one of the world's leading MNOs (as detailed in Section \ref{sec:PerformanceEvaluation}). 

\textbf{Optimization Problem}. The profit ($\Pi$) maximization considers the revenue obtained from the admission of URLLC flows to be routed throughout the network with SLA $a(d_{SLA}^{urllc},\mu_{SLA}^{urllc})$ given that eMBB traffic already in the network works as background traffic demanding a minimum aggregated transmission rate $\mu_{SLA}^{embb}$. Costs depend on the number of VNFs deployed in each pair vDU/CU. The decision variables of the optimization problem are: 
\begin{itemize}
    \item $\phi^{s,u}_v$ -- percentage of share of the transmission rate per NSI, vDU, and transport node;
    \item $F^{urllc,u}$ -- number of admitted URLLC flows per vDU;
    \item $p^{s,u}_l$ -- NSI routing path in TN per NSI and vDU;
    \item $x^u_g$ -- choice on deploying VNF $g$ at vDU $u$ ($x^u_g=0$) or at CU ($x^u_g=1$).
\end{itemize}

The objective function of the problem is given by:
\begin{equation}
    \max \Pi,
\end{equation}
\noindent where,
\begin{equation}
    \Pi=\Psi_\mathscr{R}-\Psi_\mathscr{C},
\end{equation}
and \noindent $\Psi_\mathscr{R}$ and $\Psi_\mathscr{C}$ are calculated as follows:
\begin{equation}
\Psi_\mathscr{R}=\mathscr{R}\sum_{u\in \mathcal{U}}F^{urllc,u},
\end{equation}
\noindent where, $F^{urllc,u}$ is the number of admitted URLLC UEs at vDU $u$, and
\begin{equation}
\Psi_\mathscr{C}=\sum_{u\in \mathcal{U}}\sum_{g\in \mathcal{G}}(1-x_g^u)\mathscr{C}^{DU}_g+x_g^u\mathscr{C}^{CU}_g,
\end{equation}
\noindent where (\ref{eq:eta}) holds.
Next, we present the constraints of the problem. Initially, we assure that the total resource share is limited to 100\%:
\begin{equation}
    \sum_{s\in \mathcal{S}}\sum_{u\in \mathcal{U}}\phi^{s,u}_{v} \leq 1, \forall v\in \mathcal{V}.
    \label{eq:const-rate-1}
\end{equation}


The VNF placement must respect the processing chain defined by the radio protocol stack (see Table \ref{tab:placement}), i.e. VNF $g \in \mathcal{G}$ will only be placed at the CU if the VNF $g-1$ is already deployed at the CU:

\begin{equation}
    x^{u}_{g} \leq x^{u}_{g-1},  \forall g> 1, \forall u \in \mathcal{U}.
    \label{eq:chain-constraint}
\end{equation}


The total throughput demanded by an NSI $s$ related to vDU $u$ must not exceed the capacity allocated to the NSI at any transport node:
\begin{equation}
\sum^{F^{s,u}}_{i=1}\rho^{s,u}_i\leq R^{s,u}_v, \forall v\in\mathcal{V},\forall s\in\mathcal{S},\forall u\in\mathcal{U},
\label{eq:const-rate-2}
\end{equation}
\noindent where $\rho^{s,u}_i$ is the arrival rate of the $i$-th UE associated to the $s$-th NSI and $u$-th vDU.

The total throughput allocated to all slices at the $v$-th TN must satisfy the minimum capacity requirement $\mu^{O_j}$ associated with the functional split option $O_j$ deployed at the $u$-th vDU, as follows: 

\begin{equation}
\sum^{}_{ s \in \mathcal{S}}R^{s,u}_v\geq \mu^{O_j}, \forall v\in\mathcal{V},\forall u\in\mathcal{U}.
\label{eq:rate-req-split}
\end{equation}
For admitted UEs, the throughput demand of every UE must be satisfied according to the SLA of the associated slice:
\begin{equation}
    \rho^{s,u}_i \geq \mu^{s}_{SLA}, \forall i \in F^{s,u},\forall s\in \mathcal{S},\forall u\in \mathcal{U}.
    \label{eq:const-rate-3}
\end{equation}
The end-to-end delay for the $i$-th network flow related to the $s$-th NSI and $u$-th vDU routed through the $l$-th path can be generically written as $d_{i,l}^{s,u}$. For simplicity, hereafter, we omit $l$ in notation, designating the end-to-end delay for the $i$-th UE as $d_{i}^{s,u}$, which must satisfy the minimum latency requirement $d^{O_j}$ per split option $O_j$ deployed at the $u$-th vDU and the SLA of the URLLC service $d_{SLA}^{urllc}$, as follows:
\begin{equation}
    d_{i}^{s,u} \leq d^{O_j}, \forall i \in F^{s,u},\forall s \in  \mathcal{S}, \forall u \in  \mathcal{U},
    \label{eq:delay-req-split}
\end{equation}
\begin{equation}
    d_{i}^{urllc,u} \leq d_{SLA}^{urllc}, \forall i \in F^{urllc,u}, \forall u \in  \mathcal{U},
    \label{eq:const-delay-1}
\end{equation}
\noindent where
\begin{equation}
d_{i}^{urllc,u}=d_{i,TN}^{urllc,u}+d_{i,RAN}^{urllc,u},
\label{eq:const-delay-2}
\end{equation}
\begin{equation}
d_{i,TN}^{urllc,u}=d_{i,net}^{urllc,u}+d_{i,PD}^{urllc,u},
\label{eq:const-delay-3}
\end{equation}
\begin{equation}
d_{i,RAN}^{urllc,u}=d_{i,DU}^{urllc,u}+d_{i,CU}^{urllc,u}.
\label{eq:const-delay-4}
\end{equation}
From (\ref{eq:const-delay-2}) to (\ref{eq:const-delay-4}), we state the end-to-end delay as the sum of transport delays throughout TN and processing delays in the RAN nodes. The former is the sum of the queueing delay estimated using NC ($net$) and the \ac{PD}. The latter is the sum of the processing delays related to the VNFs deployed at the DU and CU. In the following subsection, we introduce closed-form equations for such estimates.

\subsection{End-to-End Network Delay}

We use NC to propose a closed-form equation to calculate \textbf{Queueing Delay} ($d^{urllc,u}_{i,net}$) in our network slicing scenario. After, we present simple approaches to estimate \textbf{Processing Delay} ($d_{i,RAN}^{urllc,u}$)  and \textbf{Propagation Delay} ($d_{i,PD}^{urllc,u}$). For clarity, we define processing delay as the time RAN nodes (a DU and a CU) spend --- where VNFs are deployed --- to process incoming data. We define transmitting delay as the time a transport node spends to send data toward the next transport node. The transmitting delay is already considered in the queueing delay expression of Theorem \ref{theo:proposition}. As stated in (\ref{eq:const-delay-2}), the end-to-end network delay is the sum of queueing, processing, and propagation delays.
\hfill\break

\noindent 
\textbf{Queueing Delay.} We present Theorem \ref{theo:proposition} using \ac{NC} to estimate an upper-bound for queueing delay for the \ac{FoI} $f_{0}$ competing for transmission resources with other flows throughout a tandem system composed of $V$ transport nodes. After, we propose Remarks \ref{rem:remark1} and \ref{rem:remark2}. The former computes queueing delay to slice $s$ and vDU $u$ based on the \ac{GPS} scheduling algorithm. The latter extends results from Remark \ref{rem:remark1}, obtaining an upper-bound for queueing delay for a more comprehensive scenario with tree-based topologies with distinct routing paths.


\begin{theorem}
\label{theo:proposition}
Let $f_{0}$ be a \ac{FoI} competing for resources with all other $f_i$ flows,$\forall i\neq 0$, through a network path composed by $V$ FIFO servers in a feed-forward network. A leaky-bucket process bounds the arrival curve to a FIFO server with rate and burst denoted by $\rho$ and $\sigma$. Moreover, a server service curve is a rate-latency function with rate and fixed latency denoted by $R$ and $T$. Therefore, an upper-bound for the transport network delay for flow $f_{0}$ is calculated as follows:
\begin{multline}
	d_{f_0,net}=\sum_{v=0}^{V-1}T_{v_v}+\frac{\sigma_{f_0,v_0}}{R_{net}-\rho_{y(f_0)}}+\\
	+\sum_{v=0}^{V-1}\left(\frac{\rho_{y(f_0)}\sum_{j=0}^{k-1}T_{v_j}+\sigma_{y(f_0),v_0}}{R_{v_v}}\right).
	\label{eq:de2e_3}
\end{multline}


\end{theorem}


\renewcommand\qedsymbol{$\blacksquare$}

\begin{proof}
Consider a transport node transmitting input traffic flows $f_i$, where $i\in \{0,1\}$, i.e., $f_0$ and $f_1$, which are bound by the following leaky-bucket arrival curves $\alpha_{f_0}(t)$ and $\alpha_{f_1}(t)$. This transport node offers a service curve $\beta_p(t)$ with constant transmission capacity $R$ and fixed latency $T$, i.e., a rate-latency function. For completeness, $\alpha(t)$ and $\beta_p(t)$ are written in the following, where (and henceforth) the argument $t$ and index $p$ are suppressed to simplify notations:
\begin{equation}
	\alpha=\rho t + \sigma,
\end{equation}
\begin{equation}
	\beta=R[t-T]^+.
\end{equation}
In the \ac{DNC}, the \ac{SFA} with \ac{FIFO} multiplexing \cite{zhou2020survey} can calculate the upper delay bound. In this case, we calculate $t$ to compute the delay of flow $f_0$, for which the following equation holds:
\begin{equation}
	\beta ^{l.o.f_0}=\sigma_{f_0},
\end{equation}
\noindent where $\beta^{l.o.}$ stands for the leftover service curve, defined in NC as the minimum level of service that a flow receives when more than one flow passes the same transport node \cite{zhou2020survey}.
Therefore, the leftover service curve is calculated as follows \cite{fidler2010survey}:
\begin{equation}
	\beta ^{l.o.f_0}=\beta \ominus\alpha_{f_1}=\beta_{R^{l.o.f_0},T^{l.o.f_0}}.
 \label{eq:betalo}
\end{equation}
Subsequently, we calculate $R^{l.o.f_0}$ and $T^{l.o.f_0}$, as follows:
\begin{equation}
	R^{l.o.f_0}=R-\rho_{f_1}.
  \label{eq:Rlo}
\end{equation}
The value for $T^{l.o.f_0}$ is obtained calculating $t$ to hold $\beta=\sigma_{f_1}$, as follows:
\begin{equation}	T^{l.o.f_0}=T+\frac{\sigma_{f_1}}{R}.
 \label{eq:Tlo}
\end{equation}
Using (\ref{eq:Tlo}) and (\ref{eq:Rlo}) in (\ref{eq:betalo}), we have:
\begin{equation}
	\beta^{l.o.f_0}=(R-\rho_{f_1})[t-\frac{\sigma_{f_1}}{R}-T]^+,
  \label{eq:beta_s0}
\end{equation}
\begin{equation}
	(R-\rho_{f_1})[t-\frac{\sigma_{f_1}}{R}-T]^+=\sigma_{f_0},
\end{equation}
\begin{equation}
	t=\frac{\sigma_{f_0}+\sigma_{f_1}}{R-\rho_{f_1}}+T-\frac{\rho_{f_1}\sigma_{f_1}}{R(R-\rho_{f_1})}.
\end{equation}
After simplification, we obtain:
\begin{equation}
t=T+\frac{\sigma_{f_1}}{R}+\frac{\sigma_{f_0}}{R-\rho_{f_1}}.
\end{equation}
To facilitate understanding, we name the transport node mentioned above as $v_0$ and we assume its interconnection with node $v_1$ forming a tandem system with two transport nodes ($v_0$ and $v_1$). Afterward, we generalize the obtained results for tandem systems with $V$ nodes. Therefore, we also calculate the leftover service curve  for $v_1$ as follows:
\begin{equation}
\beta_{v_1}^{l.o.f_0}=\beta_{v_1}\ominus \alpha_{f_0,v_1},
\end{equation}
\noindent where $\alpha_{f_i,v_v}$ stands for the arrival curve of the $i$-th flow in the $v$-th transport node.

Note that due to the burstiness increase \cite{schmitt2008delay}, the arrival curve in $v_1$ differs from that in $v_0$, i.e.,
\begin{equation}	\alpha_{f_0,v_1}=\alpha_{f_0,v_0} \oslash \beta_{v_0}^{l.o.f_0}.
\end{equation}
Subsequently, we calculate $\rho_{f_0,v_1}$ and $\sigma_{f_0,v_1}$:
\begin{equation}
\rho_{f_0,v_1}=\rho_{f_0,v_0},
\end{equation}
\begin{equation}
	\sigma_{f_0,v_1}=\alpha_{f_0,v_0}(T_{v_0})=\rho_{f_0,v_0}T_{v_0}+\sigma_{f_0,v_0}.
\end{equation}
Similarly, we obtain $\rho_{f_1,v_1}$ and $\sigma_{f_1,v_1}$ to obtain the leftover service curve for $v_1$, as follows:
\begin{equation}
\rho_{f_1,v_1}=\rho_{f_1,v_0},
\end{equation}
\begin{equation}
	\sigma_{f_1,v_1}=\alpha_{f_1,v_0}(T_{v_0})=\rho_{f_1,v_0}T_{v_0}+\sigma_{f_1,v_0},
\end{equation}
\begin{equation}
	R_{v_1}^{l.o.f_0}=R_{v_1}-\rho_{f_1,v_1},
\end{equation}
\begin{equation}
	T_{v_1}^{l.o.f_0} \rightarrow R_{v_1}[t-T_{v_1}]^+=\sigma_{f_1,v_1},
\end{equation}
\begin{equation}	T_{v_1}^{l.o.f_0}=T_{v_1}+\frac{\sigma_{f_1,v_0}}{R_{v_1}}+\frac{\rho_{f_1,v_0}T_{v_0}}{R_{v_1}},
\end{equation}
\begin{equation}
	\beta_{v_1}^{l.o.f_0}=(R_{v_1}-\rho_{f_1,v_1})[t- T_{v_1}-\frac{\sigma_{f_1,v_0}}{R_{v_1}}-\frac{\rho_{f_1,v_0}T_{v_0}}{R_{v_1}}]^+.
 \label{eq:beta_s1}
\end{equation}

After the leftover service curves for transport nodes ($v_0$ and $v_1$) have been calculated for flow $f_0$, the network leftover service curve ($\beta^{l.o.f_0}_{net}$) is computed using the concatenation property stated in (\ref{eq:conc}), that is:
\begin{equation}
\beta^{l.o.f_0}_{net}=\beta^{l.o.f_0}_{v_0}\otimes \beta^{l.o.f_0}_{v_1},
 \label{eq:conv_s0_s1}
\end{equation}
\begin{equation}
\beta^{l.o.f_0}_{net}=\inf_{0\leq \tau \leq t}[{\beta^{l.o.f_0}_{v_0}(t-\tau)+\beta^{l.o.f_0}_{v_1}(\tau)}],
\end{equation}
\begin{equation}\beta^{l.o.f_0}_{net}=\beta_{min(R_{v_0}^{l.o.f_0},R_{v_1}^{l.o.f_0}),(T_{v_0}^{l.o.f_0}+T_{v_1}^{l.o.f_0})}.
\end{equation}

The delay is the value of $t$ that fulfills the following: 
\begin{equation}
\beta^{l.o.f_0}_{net}=\sigma_{f_0,v_0}.
\label{eq:leftover_e2e_2hops}
\end{equation}

Replacing (\ref{eq:beta_s0}) and (\ref{eq:beta_s1}) in (\ref{eq:conv_s0_s1}), we calculate $t$ that fulfills (\ref{eq:leftover_e2e_2hops}) as follows:
\begin{multline}
t=T_{v_0}+T_{v_1}+\frac{\sigma_{f_0,v_0}}{\min(R_{v_0}-\rho_{f_1,v_0},R_{v_1}-\rho_{f_1,v_1})}+\frac{\sigma_{f_1,v_0}}{R_{v_0}}+\\
        +\frac{\rho_{f_1}T_{v_0}+\sigma_{f_1,v_0}}{R_{v_1}}.
        \label{eq:delay_2servers}
\end{multline}

A network delay bound $(d_{f_0,net})$ for flow $f_0$ is a direct generalization of (\ref{eq:delay_2servers}) for feed-forward networks with $V$ servers and $F$ network flows as follows:
\begin{multline}
	d_{f_0,net}=\sum_{v=0}^{V-1}T_{v_v}+\frac{\sigma_{f_0,v_0}}{R_{net}-\sum_{i=1}^{F-1}\rho_{f_i}}+\frac{\sum_{i=1}^{F-1}\sigma_{f_i,v_0}}{R_{v_0}}+\\
	+\sum_{v=1}^{V-1}\left(\frac{\sum_{i=1}^{F-1}\rho_{f_i}\sum_{j=0}^{v-1}T_{v_j}+\sum_{i=1}^{F-1}\sigma_{f_i,v_0}}{R_{v_v}}\right),
	\label{eq:de2e_1}
\end{multline}
\noindent where,
\begin{equation}
	R_{net} = \min_{\forall v}(R_{v_v}),
\end{equation}
\begin{equation}
    F_{v_v}=F, \forall v \in \mathcal{V},
    \label{eq:relax_1}
\end{equation}
\begin{equation}
    \rho_{f_i,v_v}=\rho_{f_i}, \forall v \in \mathcal{V},
\end{equation}
\noindent
and the following constraint is fulfilled:
\begin{equation}
    \sum_{i=0}^{F-1}\rho_{f_i} \leq R_{net}.
\end{equation}

To simplify notation in (\ref{eq:de2e_1}), let $f_0$ be the \ac{FoI} (\textit{through-traffic}) and $y(f_0)$ the function that represents the amount of all other flows (\textit{cross-traffic}), implying that:
\begin{equation}
    \rho_{y(f_0)}=\sum_{i=1}^{F-1}\rho_{f_i},
\end{equation}
\begin{equation}
    \sigma_{y(f_0)}=\sum_{i=1}^{F-1}\sigma_{f_i}.
\end{equation}

Therefore, (\ref{eq:de2e_1}) can be rewritten as:
\begin{multline}
	d_{f_0,net}=\sum_{v=0}^{V-1}T_{v_v}+\frac{\sigma_{f_0,v_0}}{R_{net}-\rho_{y(f_0)}}+\\
	+\sum_{v=0}^{V-1}\left(\frac{\rho_{y(f_0)}\sum_{j=0}^{v-1}T_{v_j}+\sigma_{y(f_0),v_0}}{R_{v_v}}\right).
	\label{eq:de2e_2}
\end{multline}

\end{proof}


\noindent
\begin{remark}
\label{rem:remark1}
We leverage the network slicing concept to include in (\ref{eq:de2e_2}) the resource allocation to distinct NSIs at each transport node. In (\ref{eq:de2e_2}), the number of flows in the tandem system is constant throughout the network, and $R_{net}$ is the network bottleneck. However, for different NSIs, we can have bottlenecks at different transport network nodes. Therefore, we extend (\ref{eq:de2e_2}) considering a specific number of NSIs and cross-traffic flows at each transport node, relaxing condition (\ref{eq:relax_1}) for a more general purpose. In this order, we include $s$ and $u$ to index flows based on their NSIs and vDUs. We also assume that resources of the $v$-th transport node can be shared among distinct NSIs using \ac{GPS} according to (\ref{eq:weights}) and (\ref{eq:phi}), as follows:
\begin{multline}
	d_{f_0,net}^{s,u}=\sum_{v=0}^{V-1}T_{v_v}+\frac{\sigma_{f_0,v_0}^{s,u}}{\min_{0\leq v \leq V-1}{\{R_{v_v}-\rho_{y(f_0),v_v}^{s,u}\}}}+\\
	+\sum_{v=0}^{V-1}\left(\frac{\sigma_{y(f_0),v_v}^{s,u}+\sum_{i=1}^{F_{v_v}^{s,u}-1}\left(\sum_{j=0}^{v-1}\rho_{f_i,v_j}^{s,u}T_{v_j}\right)}{R_{v_v}^{s,u}}\right).
	\label{eq:de2e_4}
\end{multline}

\noindent
where 
\begin{equation}
    F_{v_v}^{s,u}\leq F_{v_v}, \forall v \in \mathcal{V}, \forall s \in \mathcal{S}, \forall u \in \mathcal{U}.
    \label{eq:relax_2}
\end{equation}
\end{remark}

\noindent
\begin{remark}
\label{rem:remark2}
We extend (\ref{eq:de2e_4}) to comprehend a tree-based network topology, where network traffic flow $f^{s,u}_i$ is routed through a certain number of transport nodes before entering in the first transport node belonging to the path $\mathcal{P}_l^{s,u}$ of the FoI $f^{s,u}_0$, where ${}^{*}\sigma_{f_i,v_v}$ denotes the burstiness increase for flow $f^{s,u}_i$ in the $v$-th transport node, where $v_v\in \mathcal{P}^{s,u}_l$. The value of ${}^{*}\sigma_{f_i,v_v}$ is calculated in function of $H_{f_i,v_v}$, the number of transport nodes $f^{s,u}_i$ was routed before reaching transport node $v$, obtaining the network delay $d_{f_0,net}^{s,u}$ in the \ac{TN}, as follows:
\begin{multline}
d_{f_0,net}^{s,u}=\frac{\sigma_{f_0}^{s,u}}{\min_{v_v\in\mathcal{P}_l^{s,u}}{\{R_{v_v}^{s,u}-\rho_{y(f_0),v_v}^{s,u}\}}}+\\
     +\sum_{v_v \in \mathcal{P}_l^{s,u}}\left(T_{v_v}+\frac{\sum_{f_i\in \mathcal{F}_{v}^{s,u},i\neq 0}{}^*\sigma_{f_i,v_v}^{s,u}}{R_{v_v}^{s,u}}\right),
	\label{eq:de2e_app}   
\end{multline}

\noindent where
\begin{equation}    {}^{*}\sigma_{f_i,v_v}^{s,u}=\sigma_{f_i}^{s,u}+\rho_{f_i}^{s,u}\sum_{h\in\mathcal{H}_{f_i,v_v}}T_{h},
    \label{burst3}
\end{equation}

\noindent and $\mathcal{H}_{f_i,v_v}$ is the set of transport nodes through which flow $f_i$ crossed before entering the $v$-th transport node.
\hfill\break
  
\end{remark}

Section \ref{sec:PerformanceEvaluation} uses Remark \ref{rem:remark2} to estimate $d^{urllc,u}_{i,net}$ in (\ref{eq:const-delay-3}), i.e., obtaining network delay for the $i$-th UE of URLLC NSI with traffic routed to the $u$-th vDU . 
\hfill\break

\noindent
\textbf{Processing Delay.} For the following analyses, the unit for processing capacity is the \ac{RC}. An RC is equivalent to the processing capacity of one core of a predefined multicore processor.
An RC can process the full \ac{NG-RAN} stack for a flow of $X$ Gbps in an interval of $Z$ seconds. 

In Table \ref{tab:placement}, we observe the VNF placement at the RU, DU, and CU according to the split option. The VNFs are identified by the names of each layer of the protocol stack, except for the PHY layer, which is subdivided into three sublayers designated in Table \ref{tab:placement} as PHY-A, B, and C. While PHY-A runs RF processes and \ac{FFT}, PHY-B and C run mainly mapping, and modulation \& encoding procedures, respectively. The processing delay varies with the number of VNFs deployed at the DU and CU. Each VNF has its specific processing demand. Then, let $z_g$ be the processing time percentage of the reference value $Z$ from the $g$-th VNF. For the DU and CU processing capacities, we assume that the processing time linearly turns down with the number of RCs, as long it is possible to parallelize as many threads as the number of RCs, which is simplified, although reasonable, the assumption for a small number of RCs. Therefore, the processing delay is estimated as follows: 
\begin{equation}
d_{RAN}^{s,u}=\frac{Z}{X}\left(\sum_{i=0}^{F^{s,u}-1}\rho_i^{s,u}\right)\sum_{g\in \mathcal{G}}\left(\frac{x_g^u z_g}{K^u} + \frac{(1-x_g^u) z_g}{K_0}\right) ,
\end{equation}
\noindent where $K^u$ is the number of RCs at the $u$-th vDU, while $K_0$ is the number of RCs at the CU.

\begin{table}[htb]
    \centering
\caption{Placement of RAN VNFs per split option}
    \label{tab:placement}
    \setlength{\tabcolsep}{4pt}
\begin{tabular}{|c|c|c|c|c|c|c|c|}
\hline
\textbf{Split} & \textbf{PHY-A}& \textbf{PHY-B}& \textbf{PHY-C}&\textbf{MAC}&\textbf{RLC}&\textbf{PDCP}&\textbf{RRC}\\
\hline
\hline
$O_1$ & RU & DU \cellcolor{Gray}& DU \cellcolor{Gray}& DU \cellcolor{Gray}& DU \cellcolor{Gray}& DU \cellcolor{Gray}& CU \cellcolor{LightCyan}\\
\hline
$O_2$  & RU & DU \cellcolor{Gray}& DU \cellcolor{Gray}& DU \cellcolor{Gray}& DU \cellcolor{Gray}& CU\cellcolor{LightCyan} & CU\cellcolor{LightCyan}\\
\hline
$O_4$  & RU & DU \cellcolor{Gray}& DU \cellcolor{Gray}& DU \cellcolor{Gray}& CU \cellcolor{LightCyan}& CU \cellcolor{LightCyan}& CU\cellcolor{LightCyan}\\
\hline
$O_6$  & RU & DU \cellcolor{Gray}& DU \cellcolor{Gray}&  CU \cellcolor{LightCyan}& CU \cellcolor{LightCyan}& CU \cellcolor{LightCyan}& CU\cellcolor{LightCyan}\\
\hline
$O_8$  & RU & DU \cellcolor{Gray}& CU \cellcolor{LightCyan}&  CU \cellcolor{LightCyan}& CU \cellcolor{LightCyan}& CU \cellcolor{LightCyan}& CU\cellcolor{LightCyan}\\
 \hline
$O_9$  & RU & CU \cellcolor{LightCyan}& CU \cellcolor{LightCyan}& CU \cellcolor{LightCyan}& CU \cellcolor{LightCyan}& CU \cellcolor{LightCyan}& CU\cellcolor{LightCyan}\\
\hline
\end{tabular}
\end{table}

\hfill\break

\noindent
\textbf{Propagation Delay.} Let $||v_{j+1}-v_{j}||$ be the distance between two consecutive transport nodes (internode distance) in the route $\mathcal{P}_{l}^{s,u}$ composed of $V_l^{s,u}$ nodes. Assuming $c$ as the light speed, we estimate the end-to-end PD for a network flow as follows: 
\begin{equation}
d_{i,PD}^{s,u}=\frac{1}{c}\sum_{j=0}^{V_l^{s,u}-2}  ||v_{j+1}-v_{j}|| . 
\label{eq:pd}
\end{equation}
Note that (\ref{eq:pd}) does not depend on $i$, i.e., the PD is equal for all flows in the $s$-th NSI related to the $v$-th vDU.


\section{Performance 
Evaluation}\label{sec:PerformanceEvaluation}
We evaluate in this section our proposal for resource allocation to slicing-based networks with time-sensitive applications, such as the URLLC use case. For such a purpose, we first introduce the employed methodology. We also detail the setup of our simulations and we briefly describe the software tools used in our simulations and the assumptions we have drawn during the simulation design and operation. Finally, we show the obtained results, comparing them with benchmarks and providing discussions and insights.

\subsection{Methodology and Parameters Setup}
To evaluate our proposals, we considered a disaggregated 5G NG-RAN network topology connecting UEs to 5GC and we employed the following tools: IBM CPLEX, DiscoDNC, and OMNeT++. We use IBM CPLEX to obtain the optimal solution for the problem formulation presented in Section \ref{sec:SystemModel}. Regarding comparisons with benchmarks and simulation results, on the one hand, we use DiscoDNC to automate the estimation of theoretical bounds on end-to-end delay, and we leverage the toolset provided by DiscoDNC to obtain results through diverse analyses provided by Network Calculus Theory. On the other hand, we also conduct simulations using OMNeT++ (event-driven simulation tool), getting end-to-end delay results in a more realistic network scenario where data packetization, buffer occupancy, and resource scheduling dynamics are considered. 

\begin{figure}[htb]
    \centering
    \includegraphics[width=0.48\textwidth]{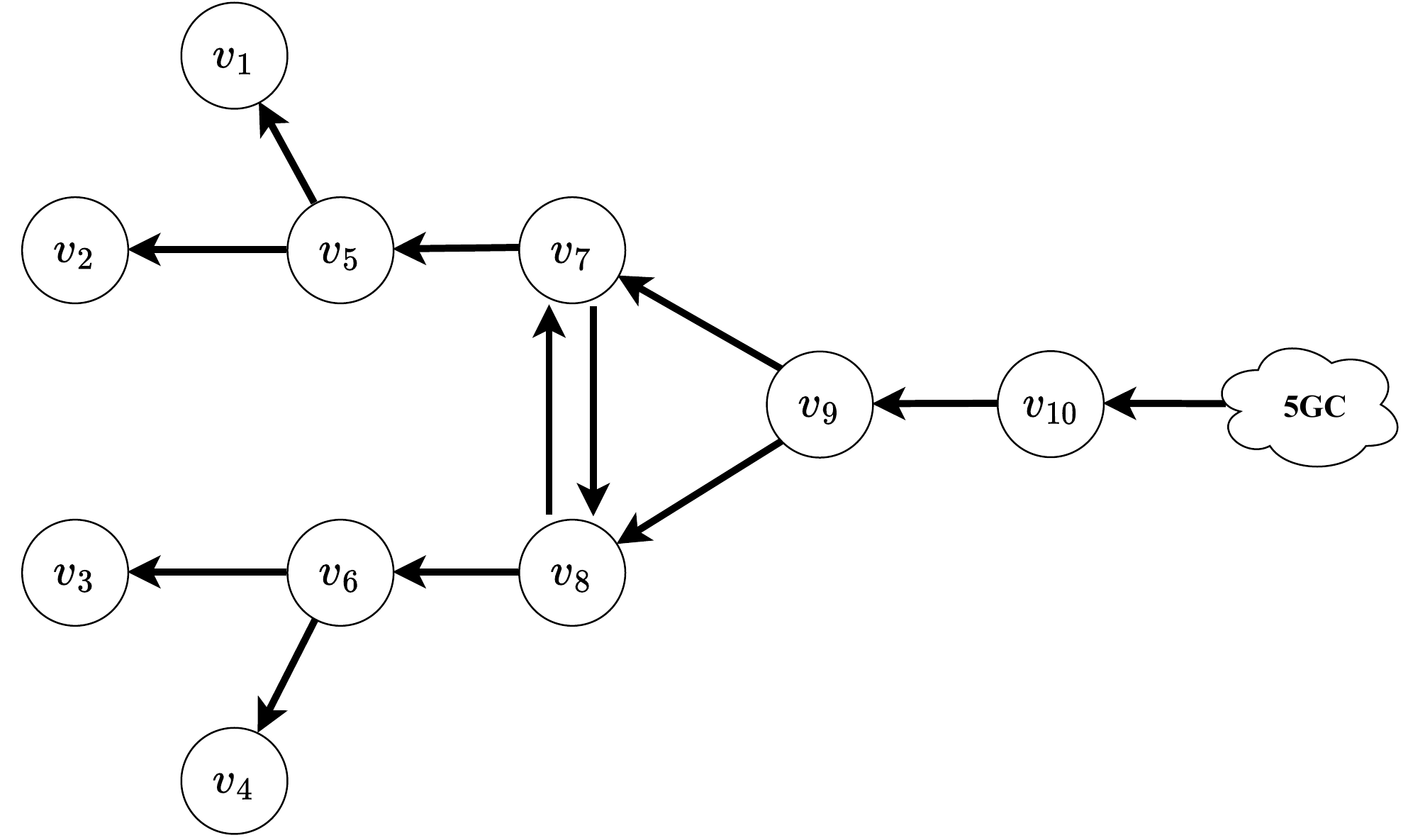}
    \caption{Transport network topology for performance evaluation.}
    \label{fig:Server-Graph-Routes}
\end{figure}

The results presented in this section were obtained through simulations in the downlink direction using the transport network topology depicted in Fig. \ref{fig:Server-Graph-Routes}. 
Our scenario is composed of a single CU ($v_{10}$), two DUs ($v_{5}$ and $v_{6}$) split into four vDUs ($v_{1}$, $v_{2}$, $v_{3}$, and $v_{4}$) co-localized with four RUs. All DUs connect to the CU through a midhaul transport network (ring composed of $v_7$, $v_8$, and $v_9$ transport nodes). Every RU supports connections to UEs using PDU sessions (flows) associated with instances of two different slicing use cases: URLLC and eMBB. NSIs of these use cases transport traffic with fixed packet sizes of 128 bytes and 1500 bytes, respectively. The arrival processes for the URLLC NSI are bounded by the arrival curve $\alpha^{urllc}$. Moreover, the eMBB NSI is considered a bandwidth consumer, and four aggregated eMBB traffic demands  are evaluated, as listed in Table \ref{Tab:eMBB}. Therefore, we name as an \textit{optimization instance} each one of the network scenarios with a specific set of eMBB demands. Our aim is to evaluate all possible scenarios with this set of varying demands. Assuming Fig.~\ref{fig:Server-Graph-Routes} as the default topology with four vDUs, we obtain 256 \textit{optimization instances} for our evaluations. 

The service characteristics for the URLLC NSI is similar to that presented in \cite{kalor2018network}, i.e., multiple robot units in factories in the context of Industry 4.0, where such units are controlled by periodic transmissions of data with a packet size of 128 bytes and inter-packet time of 1 ms. The end-to-end delay requirement for this service is 1 ms. Table \ref{tab:simulationparameters} summarizes the parameters for the arrival and service curves used in this article, the capacities for each transport node depicted in Fig. \ref{fig:Server-Graph-Routes}, and other parameter values employed in the evaluation. Our performance evaluation is carried out in a network scenario conservatively dimensioned since we intend to show its efficiency even for scenarios with capacity restrictions. Although, our approach can be applied to different topologies and network dimensioning.

\begin{table}[htb]
\centering
	\caption{Parameter Setup}
        \label{tab:simulationparameters}

	\begin{threeparttable}

	\begin{tabular}{|l|l|}
            \hline
		\textbf{Parameter} & \textbf{Value}\\
            \hline
            \hline \rowcolor{Gray}
            Bandwidth &  20 MHz\\
            \hline 
            $B$,$S$,$U$,$G$,$V$ & 4, 2, 4, 6, 10\\
            \hline \rowcolor{Gray}
            $F_{max}$ & 320 \\
            \hline
		$N_{RB}$ & 100 \\
		\hline \rowcolor{Gray}
            $SR$ & 30.72 Mbps \\
		\hline
            $N^{RB}_{SC}$ & 12  \\
		\hline \rowcolor{Gray}
            $N^{SUB}_{SYM}$ & 14  \\
		\hline
            $N_L$ & 2  \\
		\hline \rowcolor{Gray}
            $Hdr_{PDCP}$, $Hdr_{RLC}$, $Hdr_{MAC}$  & 2, 5, 2 B  \\
           \hline
            $N_{DL}^{TBS}$ & 2  \\
		\hline \rowcolor{Gray}
            $FAPI_{DL}$ & 1.5 Mbps  \\
            \hline 
            $RefSym_{REs}$ & 6  \\
		\hline\rowcolor{Gray}
            MCS & 28 \\
		\hline 
		MIMO &  2x2\\
		\hline\rowcolor{Gray}
		$\gamma$, $\eta$ &  0.118, 0.2585\\
            \hline 
		$\zeta$ &  0.5571\\
		\hline\rowcolor{Gray}
            TBS & 75376 bits \\
		\hline
		$PDCCH_{REs}$&  144\\
		\hline \rowcolor{Gray}
		$N_{IQ}$&  32 bits\\
            \hline
            $N_{AP}$&  2 \\
            \hline \rowcolor{Gray}
		$IP_{pkt}^{urllc}$&  128 B\\
		\hline
            $IP_{pkt}^{embb}$&  1500 B\\
		\hline\rowcolor{Gray}
            $d^{urllc}_{SLA}$ & 1 ms\\
            \hline 
            $\mu^{urllc}_{SLA}$ & 1.024 Mbps\\
            \hline\rowcolor{Gray}
            $\mu^{embb}_{SLA}$ & Table \ref{Tab:eMBB}\\
            \hline 
            $\rho^{urllc}$ & 1.024 Mbps\\
            \hline\rowcolor{Gray}
            $\sigma^{urllc}$ & 128 B\\
            \hline
            $RC_{vDU}$,$RC_{CU}$& 16, 32\tnote{*}\\
            \hline\rowcolor{Gray}
             Internode distance & 5 Km\\
            \hline 
            $z_g$& Table \ref{tab:processing}\\
            \hline \rowcolor{Gray}
            $Z$& 750 $\mu$s\\
            \hline 
            $X$& 1 Gbps\\
            \hline\rowcolor{Gray}
            $v_1$,$v_2$,$v_3$,$v_4$& 1.2 Gbps\\
            \hline 
            $v_5$,$v_6$ & 8 Gbps\\
            \hline\rowcolor{Gray}
            $v_7$,$v_8$,$v_9$,$v_{10}$& 20 Gbps\\
            \hline
\end{tabular}
\begin{tablenotes}
   \item[*] Number of RCs at the CU per
related vDU.  
\end{tablenotes}

 \end{threeparttable}
\end{table}

\begin{table}[htb]
	\centering
\begin{threeparttable}
	\caption{Parameters for eMBB demand.}
           \label{Tab:eMBB}
	\begin{tabular}{|c|c|}
		\hline
		\textbf{eMBB demand} & \textbf{$\mu^{embb}_{SLA}$ (Mbps)} \\
		\hline
            \hline \rowcolor{Gray}
		20\% & 29.201   \\
		\hline
            40\% & 58.243    \\
		\hline\rowcolor{Gray}
		60\% & 87.109    \\
		  \hline
    	80\% & 117.81   \\
		  \hline
	\end{tabular}
  \end{threeparttable}
\end{table}

We considered that each RU transmits data to UEs using a wireless channel interface dimensioned conservatively using \ac{NR} numerology 0 with a bandwidth of 20 MHz with total capacity for data plane provided by 100 RBs per \ac{TTI}. A single RB per TTI can serve each URLLC UE. In this approach, if the amount of eMBB traffic corresponds to, for example,  20\% of the RU capacity, 80 RBs are left to URLLC traffic. Each network node has a transmission capacity determined by a rate-latency service curve. In the ring plus sink-tree topology depicted in Fig. \ref{fig:Server-Graph-Routes}, the transport nodes are distributed equally with an internode distance of 5 km, exception for RUs which are physically close to the vDUs. The ring at the midhaul interconnects the CU to the DUs and provides different paths to the network flows in downlink direction.

The number of VNFs deployed at a vDU is related to the functional split option $O_j \in \mathcal{\hat{O}}$, where $\mathcal{\hat{O}}$ is the set of functional split options defined in Subsection~\ref{sec:bg-split} (refer to Table \ref{tab:splitchoiceSCF}). The split options $O_8$ and $O_9$ include subdivisions for the PHY layer. Therefore, it is also necessary to map the processing time demanded by functions inside the PHY layer. We considered as \acf{RC} to our analyses one core of an Intel Haswell i7-4770 3.40 GHz, which processes, during a single \ac{TTI}, packets from an aggregated data rate of $X$ = 1 Gbps in the time interval of $Z$ = 750 $\mu$s. The measures used in our analyses are based on the works in \cite{garcia2018fluidran} and \cite{chun2016performance}. Similar premises were also used in \cite{ojaghi2019sliced} and \cite{ojaghi2022impact}. In such an approach, the PHY layer is divided into three main functions: \ac{FFT}, modulation, and encoding \cite{khatibi2018modelling}. Based on the measures from \cite{chun2016performance}, 26\% of the total of 64.99\% of the PHY layer processing time is demanded by modulation. In \cite{khatibi2018modelling}, the authors showed that the encoding/decoding processing time is sensitive to the \ac{MCS} order. However, processing times for \ac{FFT} and modulation functions are less sensitive to MCS variations. Using the method proposed in \cite{khatibi2018modelling} for downlink and MCS 28, the distribution of PHY processing time for FFT, modulation, encoding, and other PHY functions (mainly mapping functions) are respectively 8.95\%, 21.58\%, 54.25\%, and 15.22\%. Table~\ref{tab:processing} shows for each protocol stack layer $g$ the VNF processing time $z_g$ in percentage terms of the full-stack processing time $Z$. In this approach, PHY-A is always placed at the RU, while all other VNFs are placed at the DU or CU. 

\begin{table}[htb]
    \centering
    \begin{threeparttable}
\caption{RAN VNF processing time as percentage of the full-stack processing time (adapted from 
 \cite{chun2016performance} and \cite{khatibi2018modelling})}
    \label{tab:processing}
\begin{tabular}{|l|l|l|l|}
\hline
\textbf{Layers} & \multicolumn{2}{c|}{$z_g$: \textbf{VNF processing time (\%)}} & $g$\\
\hline
\hline \rowcolor{Gray}
RRC & \multicolumn{2}{c|}{2.17} & 1\\
\hline
PDCP & \multicolumn{2}{c|}{18.7}& 2\\
\hline\rowcolor{Gray}
RLC & \multicolumn{2}{c|}{0.91} & 3\\
\hline
MAC & \multicolumn{2}{c|}{13.24}& 4\\
\hline
\multirow{3}{*}{PHY} & \multirow{3}{*}{64.99} &  PHY-C: Modulation \& Encoding: 49.28 \cellcolor{Gray} & 5 \cellcolor{Gray}\\ \cline{3-4} 
 & & PHY-B: Mapping functions: 9.89  & 6\\ \cline{3-4}
 & & PHY-A: FFT: 5.82  \cellcolor{Gray} & \ac{RU}\tnote{*} \cellcolor{Gray}\\ \cline{1-4}
\end{tabular}
\begin{tablenotes}
   \item[*] In compliance with O-RAN (split 7.2x) and according to Table \ref{tab:placement}, \ac{FFT} always runs at the RU.
\end{tablenotes}
\end{threeparttable}

\end{table}

Regarding overheads imposed on the network traffic packets in the TN, Table \ref{tab:multiplier} presents the overhead factors for packets of 128 bytes (URLLC) and 1500 bytes (eMBB) due to the split options. From $O_1$ to $O_6$, the overhead factor complies with the 5G NR standards for each layer \cite{ETSI138321, ETSI138322, ETSI138323}, while for $O_8$ and $O_9$ (which represent subdivisions of the PHY layer), the overhead factor is  estimated taking into account the proportional of capacity demand required for each split option. We use the equations in Table \ref{tab:splitchoiceSCF} to calculate capacities' requirements $\mu^{O_j}$ for each functional split option $O_j$, assuming an assignment of all available RBs to a single UE with a good channel quality. Furthermore, we assume near-ideal one-way latency $d^{O_j}$ for functional split options $O_j$ ($O_6$, $O_8$, and $O_9$) with centralized MAC layer featured with HARQ interleaving \cite{SCF.159.07.02}. 

\begin{table}[htb]
    \centering
\caption{Multiplier factor of URLLC (128 Bytes) and eMBB (1500 Bytes) packets per split option}
    \label{tab:multiplier}
\begin{tabular}{|c|c|c|}
\hline
\multirow{2}{*}{\textbf{Split}} & \multicolumn{2}{c|}{\textbf{Multiplier}}\\
\cline{2-3}
 & \textbf{128 Bytes} & \textbf{1500 Bytes}\\
\hline\hline\rowcolor{Gray}
$O_1$ &  1.0000 & 1.0000\\
\hline
$O_2$  & 1.0157 & 1.0014\\
\hline\rowcolor{Gray}
$O_4$  & 1.0547 & 1.0047\\
\hline
$O_6$  & 1.0704 & 1.0060\\
\hline\rowcolor{Gray}
$O_8$  &  6.6214 & 6.2235\\
 \hline
$O_9$  & 7.6338 & 7.1751\\
\hline
\end{tabular}
\end{table}

\subsection*{\textbf{Cash Flow Analysis}}

\begin{table}[htb]
	\caption{Real Data from Verizon Communications Inc.}
  \label{tab:verizon}
	\centering
 \begin{threeparttable}

	\begin{tabular}{|l|l|}
            \hline 
                    \textbf{Verizon Communications Inc.} & \textbf{Q3 2022}\\
                       \hline
                        \hline 
            \multicolumn{2}{|c|}{\textbf{Consumer, Business and Other}} \\
		\hline\rowcolor{Gray}
           Operating Revenues  & (millions)\\
            \hline
		\text{    } Service Revenues and Other  & \$27,666.00\\
		\hline
		\text{    } Wireless Equipment Revenues  &  \$6,575.00\\
		\hline
		\text{    } \textbf{Total Operating Revenues}  &  \textbf{\$34,241.00} \\
		\hline\rowcolor{Gray}
		Operating Costs&  (millions)\\
		\hline
		\text{    } Cost of Services  &  \$7,293.00\\
		\hline
            \text{    } Cost of Wireless Equipment &  \$7,308.00\\
		\hline
            \text{    } Selling, General and Administrative Cost &  \$7,422.00\\
		\hline
  \text{    } Depreciation and Amortization Cost & \$4,324.00\\
  \hline
  \text{    } \textbf{Total Operating Costs} & \textbf{\$26,347.00}\\
  \hline\hline
\multicolumn{2}{|c|}{\textbf{Consumer}}\\
            \hline\rowcolor{Gray}
            Operating Revenues & (millions)\\
            \hline
		\text{    } Service   & \$18,421.00 \\
		\hline
		\text{    } Wireless Equipment &  \$5,558.00\\
		\hline
  \text{    } Other &  \$1,861.00 \\
  \hline
  \text{    } \textbf{Total Operating Revenues} & \textbf{\$25,840.00}\\
  \hline\rowcolor{Gray}
  Operating Costs & (millions)\\
\hline
		\text{    } Cost of Services  &  \$4,566.00
\\
		\hline
\text{    } Cost of Wireless Equipment &  \$5,963.00
\\
		\hline
            \text{    } Selling, General and Administrative Cost &  \$4,730.00
\\
		\hline
  \text{    } Depreciation and Amortization Cost & \$3,232.00
\\
  \hline
  \text{    } \textbf{Total Operating Costs}
&\textbf{\$18,491.00}\\
\hline

  \hline\hline
  \multicolumn{2}{|c|}{\textbf{Business}}\\
            \hline\rowcolor{Gray}
            Operating Revenues &	(millions)\\
                        \hline
\text{    } Small and Medium Business 	& \$3,196.00\\
            \hline
\text{    } Global Enterprise 	& \$2,449.00\\
            \hline
\text{    } Public Sector and Other &	\$1,531.00\\
            \hline
\text{    } Wholesale &	\$661.00\\
            \hline
\text{    } \textbf{Total Operating Revenues} &	\textbf{\$7,837.00}\\ \hline\rowcolor{Gray}
Operating Costs &	(millions)\\ \hline
\text{    } Cost of Services &	\$2,653.00\\ \hline
\text{    } Cost of Wireless Equipment 	&\$1,344.00\\ \hline
\text{    } Selling, General and Administrative Cost &	\$2,063.00\\ \hline
\text{    } Depreciation and Amortization Cost&	\$1,079.00\\ \hline
\text{    } \textbf{Total Operating Costs} &	\textbf{\$7,139.00}\\ \hline\hline\rowcolor{Gray}
\multicolumn{2}{|c|}{\textbf{\textbf{Wireless Statistics}}}\\  
\hline
Postpaid Accounts (millions)&35.034\\ \hline
Postpaid Connections per Account &	3.430\\ \hline
Prepaid Connections (millions)&	23.076\\ \hline
Total Connections (millions)&	143.243\\ \hline
Prepaid ARPU &	\$31.18 \\ \hline
Postpaid ARPA &	\$149.82\\ \hline\rowcolor{Gray}
\multicolumn{2}{|c|}{\textbf{Estimated Wireless Revenues \& Costs}}\\  \hline 
Overall ARPU/month	& \$41.67\\ \hline
Revenue: Consumer \& Business (millions)	& \$17,904.91\\\hline
Cost: Consumer \& Business (millions)	& \$ 9,975.68\\\hline
Average Connections at BE point (millions)& 	79.807\\\hline \rowcolor{LightCyan}
$\zeta$ & \textbf{0.5571}\\ \hline
\end{tabular}
   \end{threeparttable}
\end{table}

In our optimization problem, $\zeta$ represents the relationship, in monetary units, between revenue and cost. To obtain a realistic value for $\zeta$ in compliance with Theorem \ref{theo:CFA}, we performed a cash flow analysis using data from the Financial and Operating Information (publicly available) of the third quarter of 2022 from Verizon, one of the world's leading MNOs. Table~\ref{tab:verizon} shows the most relevant information obtained from our analyses. Verizon offers communication services and products to final consumers, businesses, and public entities. In the quarter under review, Verizon recorded total operating revenues of \$34.2 billion, obtained from incomes with services and wireless equipment sales. Regarding the services revenues, service contracts with consumers corresponded to earns of approximately \$18.4 billion (of which \$15.5 billion with wireless services), and wireless services contracts with businesses summed approximately \$3.3 billion. The wireless services include postpaid and prepaid accounts. Our analyses estimated an overall \ac{ARPU}/month of \$41.67. The total operating cost recorded was \$26.3 billion, including costs with services, wireless equipment, general administrative expenditures, depreciation, and amortization. 

From the operational cost perspective, it is troublesome to distinguish operating costs from distinct company branches (e.g., mobile and fixed broadband) since they may have expenditures in common with infrastructure and administration. From our cash flow analysis, we isolated the revenues and costs related exclusively to wireless services for consumer and business clients. We estimated \$17.9 and approximately \$10 billion for the revenues and costs with wireless services, obtaining $\zeta=0.5571$. Although this value may vary for different MNOs, it maintains a \textit{plateau} (i.e., similar values in distinct quarters) for the same MNO since we considered only recurrent revenues and costs in our analysis. For the relationship between the cost to deploy and operate VNFs at the vDU or CU, we considered that running VNFs at the CU is cost-saving.
The authors in \cite{ojaghi2019sliced} use $\eta=0.017$ for operation and $\eta=0.5$ for deployment. Based on these values, we considered $\eta=0.2585$ as a single parameter to denote total deployment and operational cost for a VNF at the CU in contrast to the vDU.   




\subsection*{\textbf{Evaluation Tools}} 

We conduct all experiments in a Virtual Machine (VM) with 32 vCPUs, 256 GB RAM, and 100 GB of disk. VM is hosted in a server HPE ProLiant DL580 G7 with four Intel Xeon E7-4830 @ 2.13GHz. We used Python 3.8.10 and docplex 2.25.236 for implementing the optimization model and IBM CPLEX 22.1.1 as the optimal solver. 
We used Java OpenJDK 18.0.1 and DiscoDNC 2.4.4 to automate the estimation of bounds on end-to-end delay for the network topologies considered in this work using the following analyses: \ac{TFA}, \ac{SFA}, and \ac{PMOO}. We carried out estimations for TFA and SFA using \ac{FIFO} and \ac{ARB}. The latter is also known in the literature as blind multiplexing \cite{fidler2010survey}. We implemented the network scenario depicted in Fig. \ref{fig:queueingModel} in OMNeT++ 6.0, using the INET framework 4.4 to compare the theoretical results with those obtained from simulations. Moreover, we assumed a traffic model based on a Poisson process to capture the dynamic behavior of the traffic downloaded from CU to each UE. Based on this simple assumption, we obtain a more realistic scenario where arrivals are modeled as a statistical distribution encompassing a non-full buffer model. In such a model, each UE has a specific throughput demand, i.e., the transmission rate $\rho$ at which the UE downloads data. In our scenario, each active UE consumes a service related to a specific NSI (URLLC or eMBB). 

Each RU is associated with a specific vDU from the cell perspective, with different queues for different NSIs. Following a FIFO discipline, each queue receives traffic downloaded from CU (through the multiple transport nodes of the TN) to all the UEs related to a particular NSI and vDU. The buffer size is estimated not to present packet overflow, which is obtained using $q_{max}$ from (\ref{eq:q_max}). The total number of queues in the network depends on the number of NSIs and vDUs. Regarding transmission capacity, the vDU resources are shared between these queues accordingly to a \ac{WRR} scheduling algorithm, which is used as the real-world scheduling algorithm that approximates the ideal scheduling algorithm (\ac{GPS}). The \ac{WRR} weights are related to the transmission capacity share in each transport node in (\ref{eq:weights}). We estimate the number of \acp{RB} demanded by each UE since the throughput demand of each UE is known by assuming a fixed \ac{MCS} value and considering TTI as the baseline time unit for resource allocation. Additionally, the traffic is packetized with fixed packet sizes, and a small burst (equal to a packet size) is considered for each UE. 
More details about the evaluation environment and all source code are publicly available in the GitHub repository\footnote[2]{https://github.com/LABORA-INF-UFG/paper-FGKCJ-2023}.

\subsection{Results}

We present in this section results obtained from our optimization model and network simulations using OMNeT++. As previously described, we vary the total demand for URLLC and eMBB NSIs for each RU of our network scenario to obtain a range of results that provide insights into our analyses. We also carry out comparisons with baseline approaches (e.g., fixed functional splits, diverse NC analysis methods, and simulations with Round Robin scheduling) to assess the performance of our proposals.







\begin{figure}[ht]
    \centering
    \includegraphics[width=0.48\textwidth]{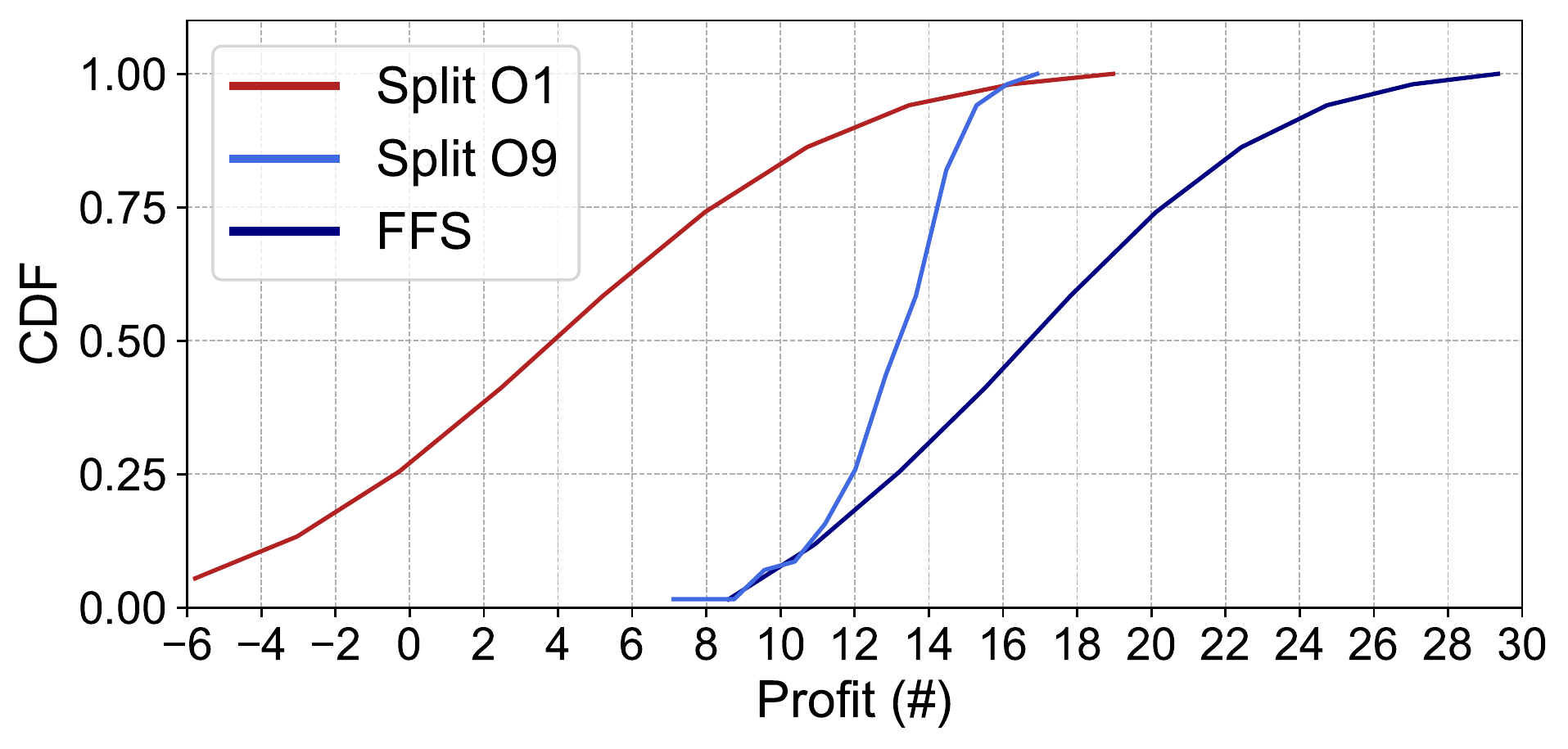}
    \caption{Profit CDF.}
    \label{fig:profitCDF}
\end{figure}

Figure \ref{fig:profitCDF} depicts the profit distribution of the solutions obtained through the 256 \textit{optimization instances} considered in our experiments. The profit value is obtained for three scenarios: fixed split $O_1$ (D-RAN), fixed split $O_9$ (C-RAN), and flexible approach for the functional split (FFS), i.e., allowing any combination of the functional splits defined in Table \ref{tab:placement}. Comparing the profit distribution for the fixed functional splits ($O_1$ and $O_9$), we observe that split $O_9$ presents a higher probability of greater profit than split $O_1$. However, the scenario using only split $O_9$, which is cost-saving compared to split $O_1$, presents solutions with a maximum profit of 17 monetary units. The scenario using only split $O_1$ achieves a maximum profit of 19 monetary units. In this context, even with a higher cost than split $O_9$, there are scenarios where split $O_1$ can admit more URLLC flows, achieving higher revenue than split $O_9$. This behavior occurs in scenarios with many URLLC flows (low eMBB demand) competing for resources, where split $O_1$ finds greater profit by admitting more flows than split $O_9$ due to its lower packet overhead. However, in scenarios with less competition for resources, i.e., with few URLLC flows in the RUs, split $O_1$ cannot balance its high cost by admitting more flows and presenting financial loss (or negative profit). This happened in nearly 25\% of the evaluated instances. 
We also observe in Fig. \ref{fig:profitCDF} that the \ac{FFS} scenario presents the best profit distribution among all evaluated instances. This distribution is due to its greater number of functional split options, allowing the optimal solver to find cost-effective solutions by choosing appropriate functional splits according to the RU demands. 


\begin{figure}[ht]
    \centering
    \includegraphics[width=0.48\textwidth]{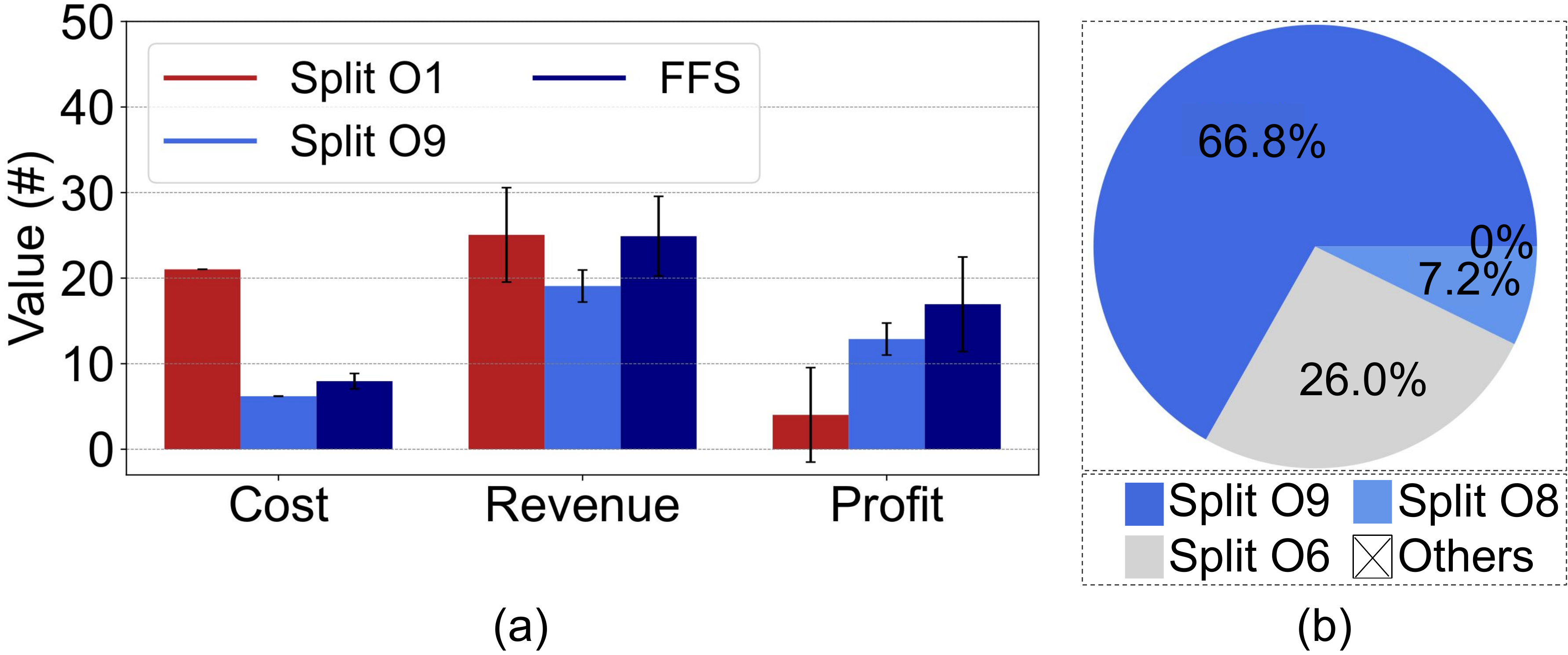}
    \caption{All scenarios.}
    \label{fig:allScenarios}
\end{figure}

Figure \ref{fig:allScenarios} presents a bar and a pie chart with an overview of the results obtained using the 256 \textit{optimization instances}. Figure \ref{fig:allScenarios} (a) shows the average cost, revenue, and profit considering split $O_1$ (D-RAN), split $O_9$ (C-RAN), and \ac{FFS} (Table \ref{tab:placement}). There is no cost variation in scenarios with fixed functional splits since the cost is modeled as a function of the VNF placement, such as split $O_1$ and split $O_9$. However, \ac{FFS} presents cost variation due to the flexibility of deploying distinct functional splits per vDU. Furthermore, split $O_1$ presents the worst case regarding the cost, as it places all VNFs (except RRC) at the vDUs. 
\ac{FFS} and split $O_1$ present higher profit than split $O_9$, despite having lower revenue. 
Even admitting fewer URLLC flows than split $O_1$, split $O_9$ compensates its low revenue with a low cost by placing VNFs at the CU. We can also observe that the best solution in terms of profit is the FFS approach, which presents costs close to the optimum (split $O_9$ (C-RAN)) and revenue close to the maximum (split $O_1$ (D-RAN)). As expected, due to its adaptability to the demand variation, the NG-RAN with flexible splits presents greater profit than that with fixed splits. 

Figure \ref{fig:allScenarios} (b) presents the choices of functional splits performed by the FFS approach. The split option $O_9$ is the most cost-saving deployment, being the most used, mainly for RUs with low demand. Another frequently used split is the split $O_6$, which presents the best cost-benefit ratio regarding the cost of VNF deployment and packet overhead, being the most used option in RUs with high demand. Finally, split $O_8$ is preferable when split $O_9$ cannot be deployed due to its stringent delay requirements. In our evaluation, the other functional split options ($O_1$, $O_2$, and $O_4$) were not selected by the optimal solver in any of the 256 \textit{optimization instances}, showing that the benefit of deploying them (less packet overhead) do not compensate the cost to deploy more VNFs at the vDUs. Naturally, this choice of splits is highly influenced by the network topology and available resources.


\begin{figure}[htb]
    \centering
    \includegraphics[width=0.45\textwidth]{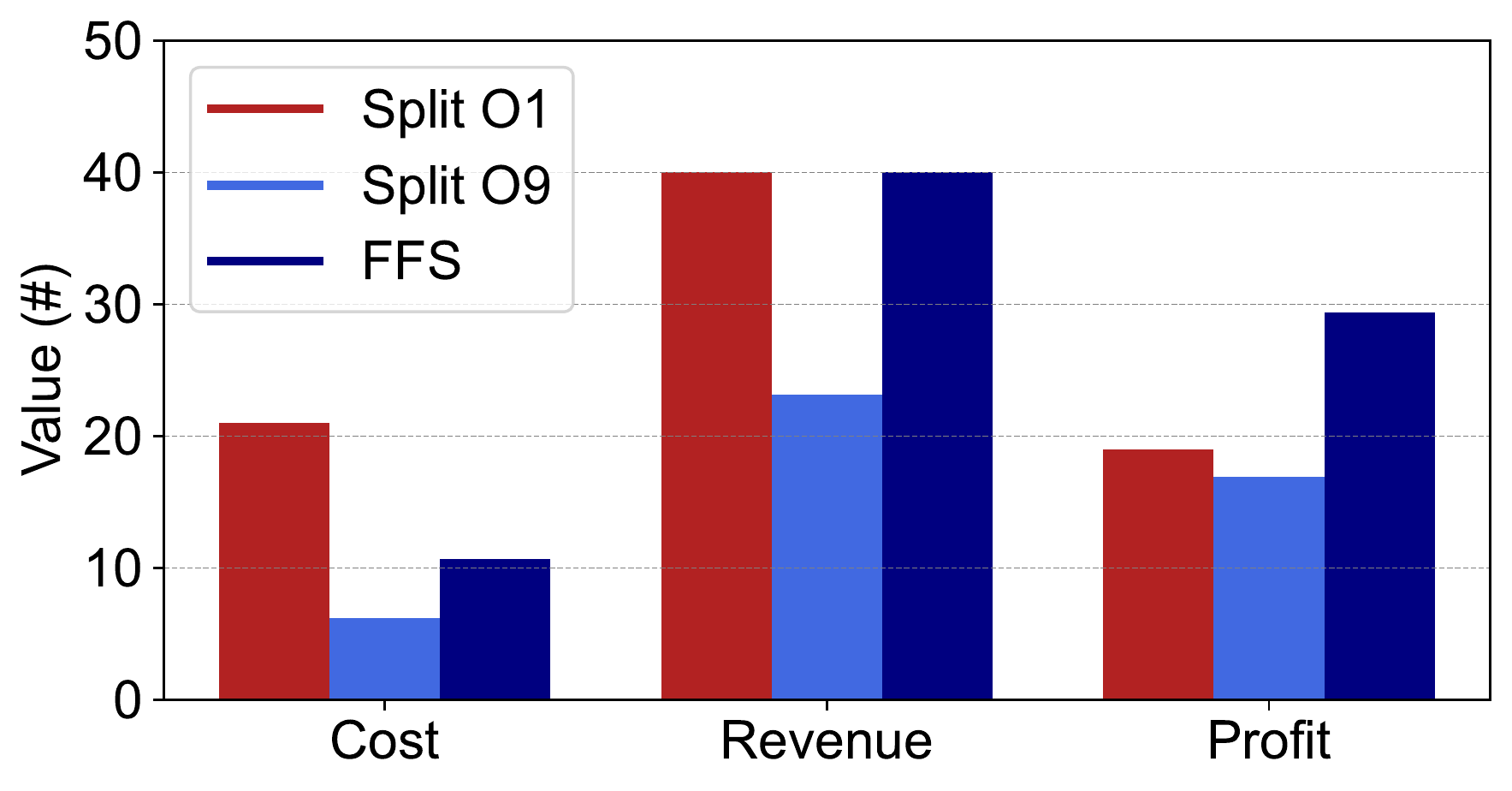}
    \caption{Scenario with 20\% of eMBB demand in each vDU.}
    \label{fig:80eMBB}
\end{figure}

In Fig. \ref{fig:80eMBB}, we explore one of the \textit{optimization instances} where split $O_1$ presented a profit greater than split $O_9$. Such a scenario is based on many URLLC flows to be admitted per RU, i.e., high resource competition. Therefore, split $O_1$ can be more resource-efficient than split $O_9$ in scenarios with resource scarcity.
In order to analyze the resource allocation results per NSI and transport node, we used the network topology depicted in Fig.~\ref{fig:Server-Graph-Routes} and data from Table \ref{tab:simulationparameters} as input to our optimization model, obtaining as results the output shown in Table \ref{tab:model_output}. In this table, we see values for $\phi^{s,u}_{v_v}$ for each transport node $v_v$ ($v_1$ to $v_{10}$), for each NSI $s$ (URLLC or eMBB) and for each vDU $u$ (1 to 4). For example, $\phi^{urllc,1}_{v_1}$ = 0.70 means that 70\% of the transmission capacity of the transport node $v_1$ is allocated to the NSI $s$ = urllc related to vDU $u$ = 1, while $\phi^{urllc,2}_{v_1}$ = 0.00 means that the NSI $s$ = urllc related to vDU $u$ = 2 does not have a route through transport node $v_1$. 

We observe, analyzing Table \ref{tab:model_output}, that all network flows followed the shortest path in the considered topology, except the eMBB traffic related to vDU $u$ = 4, which is transmitted using 10\% of the transmission capacity of transport node $v_7$. We can also observe in Table~\ref{tab:model_output} the optimal outputs regarding the function splits deployed at the vDUs and the number of URLLC UEs admitted in each RU. For vDU $u$ = 1, where $F^{urllc}$ = 80, the optimal solution deployed split $O_6$, while in all other vDUs, the optimal solution chose split $O_9$. We observe a similar solution analyzing all scenarios from Fig.~\ref{fig:profitCDF} to Fig.~\ref{fig:80eMBB}, i.e., in scenarios with a high number of UEs to be admitted in the network, splits with fewer VNFs at the CU, such as split $O_6$, is preferable since the revenue obtained with more UEs balance the high cost of more VNFs being deployed at the vDU. However, when there are fewer UEs using NSI $s$ = urllc, the cost-saving split $O_9$ becomes preferable.

\begin{table}[htb]
\centering
	\caption{Model Output Values: Optimal Solution with Objective Function equal to 14.43 monetary units.}
        \label{tab:model_output}

	\begin{threeparttable}
    \setlength{\tabcolsep}{4pt}

	\begin{tabular}{|c|c|c|c|c|}
            \hline
            $v_v$ & $\phi^{urllc,1:4}_{v_v}$& $\phi^{eMBB,1:4}_{v_v}$& Split & $F^{urllc}$ \\
            \hline
            \hline \rowcolor{Gray}
            $v_1$ &  \{0.70  0.00 0.00 0.00\} &  \{0.30 0.00 0.00 0.00\} &  $O_6$&  80\\
            \hline 
            $v_2$ &  \{0.00 0.60 0.00 0.00\} &  \{0.00 0.40 0.00 0.00\} &  $O_9$&  60\\
            \hline \rowcolor{Gray}
            $v_3$ &  \{0.00 0.00 0.50 0.00\} &  \{0.00 0.00 0.50 0.00\} &  $O_9$&  40\\
            \hline
		$v_4$ &  \{0.00 0.00 0.00 0.50\} &  \{0.00 0.00 0.00 0.50\} &  $O_9$& 20\\
		\hline \rowcolor{Gray}
            $v_5$ &  \{0.10 0.70 0.00 0.00\} &  \{0.10 0.10 0.00 0.00\} &  $-$&  $-$\\
            \hline 
            $v_6$ &  \{0.00 0.00 0.50 0.20\} &  \{0.00 0.00 0.15 0.15\} &  $-$& $-$\\
            \hline \rowcolor{Gray}
            $v_7$ &  \{0.10 0.70 0.00 0.00\} &  \{0.05 0.05 0.00 0.10\} &  $-$&  $-$\\
            \hline
		$v_8$ &  \{0.00 0.00 0.56 0.24\} &  \{0.00 0.00 0.10 0.10\} &  $-$& $-$\\
            \hline\rowcolor{Gray}
            $v_9$  & \{0.04 0.46 0.20 0.10\} & \{0.05 0.05 0.05 0.05\} & $-$ & $-$\\
            \hline 
             $v_{10}$  & \{0.04 0.48 0.22 0.06\} & \{0.05 0.05 0.05 0.05\} & $-$ & $-$\\
            \hline
      
\end{tabular}

 \end{threeparttable}
\end{table}

\begin{figure}[!ht]
    \centering
    \includegraphics[width=0.45\textwidth]{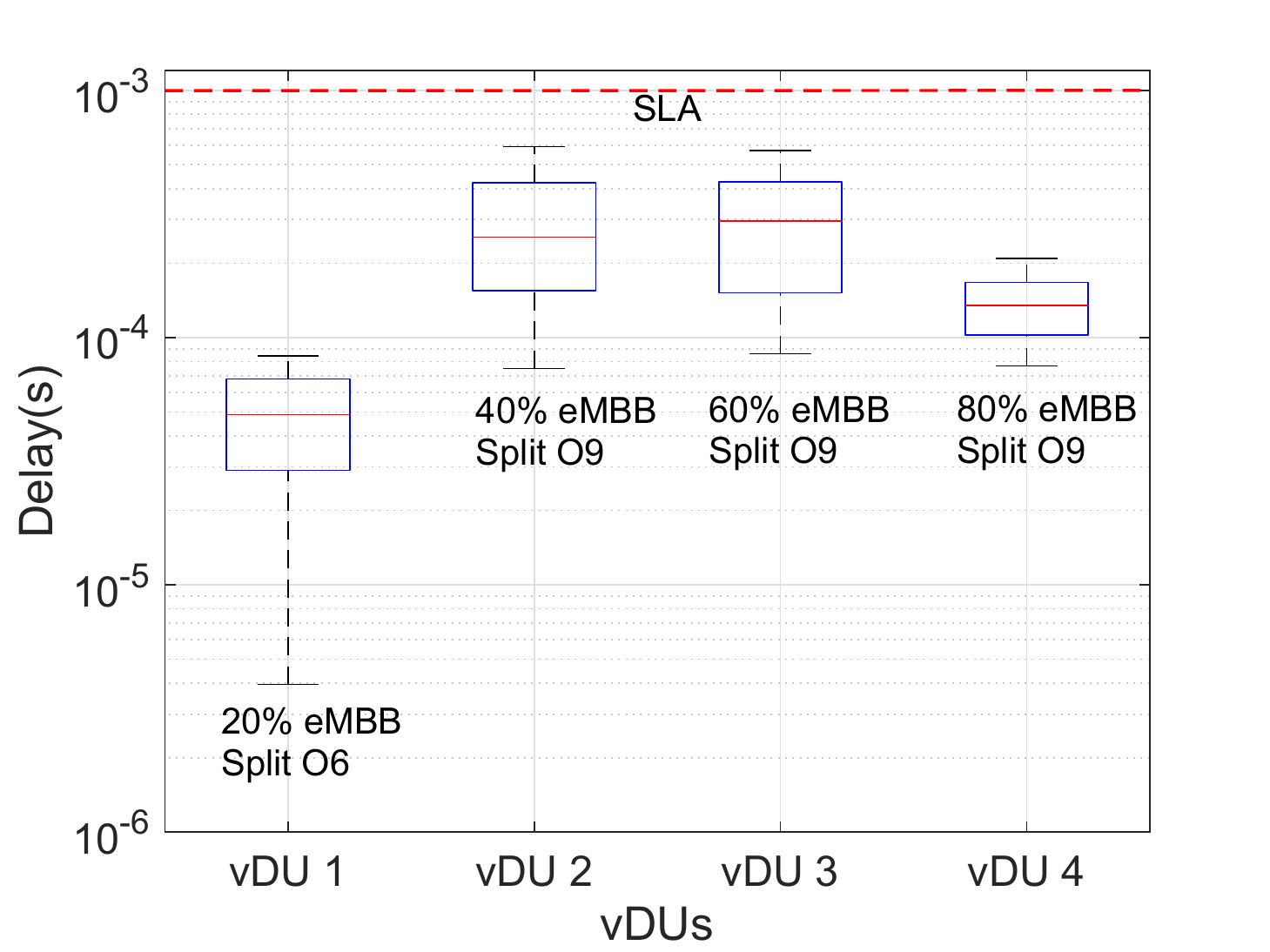}
    \caption{Boxplot and whiskers depicting the statistical distribution of maximum packet delay per vDU.}
    \label{fig:boxplot2}
\end{figure}

We used the optimal solution presented in Table \ref{tab:model_output} to set up the network scenario we have implemented in OMNeT++. Figure~\ref{fig:boxplot2} shows the boxplots and whiskers with the statistical distribution of the maximum packet delay experienced by all UEs admitted in vDUs 1 to 4. The results depicted in Fig.~\ref{fig:boxplot2} show that when split $O_9$ is chosen (vDUs 2, 3, and 4), the packet delay is greater than that observed in vDU $u$ = 1, which deploys a functional split (split $O_6$) with a smaller packet size. Moreover, the whiskers in Fig.~\ref{fig:boxplot2} show that no violations of the SLA were registered for any of the UEs, corroborating with the maximum delay guarantee provided by our proposal. 

\begin{figure}[!ht]
    \centering
    \includegraphics[width=0.45\textwidth]{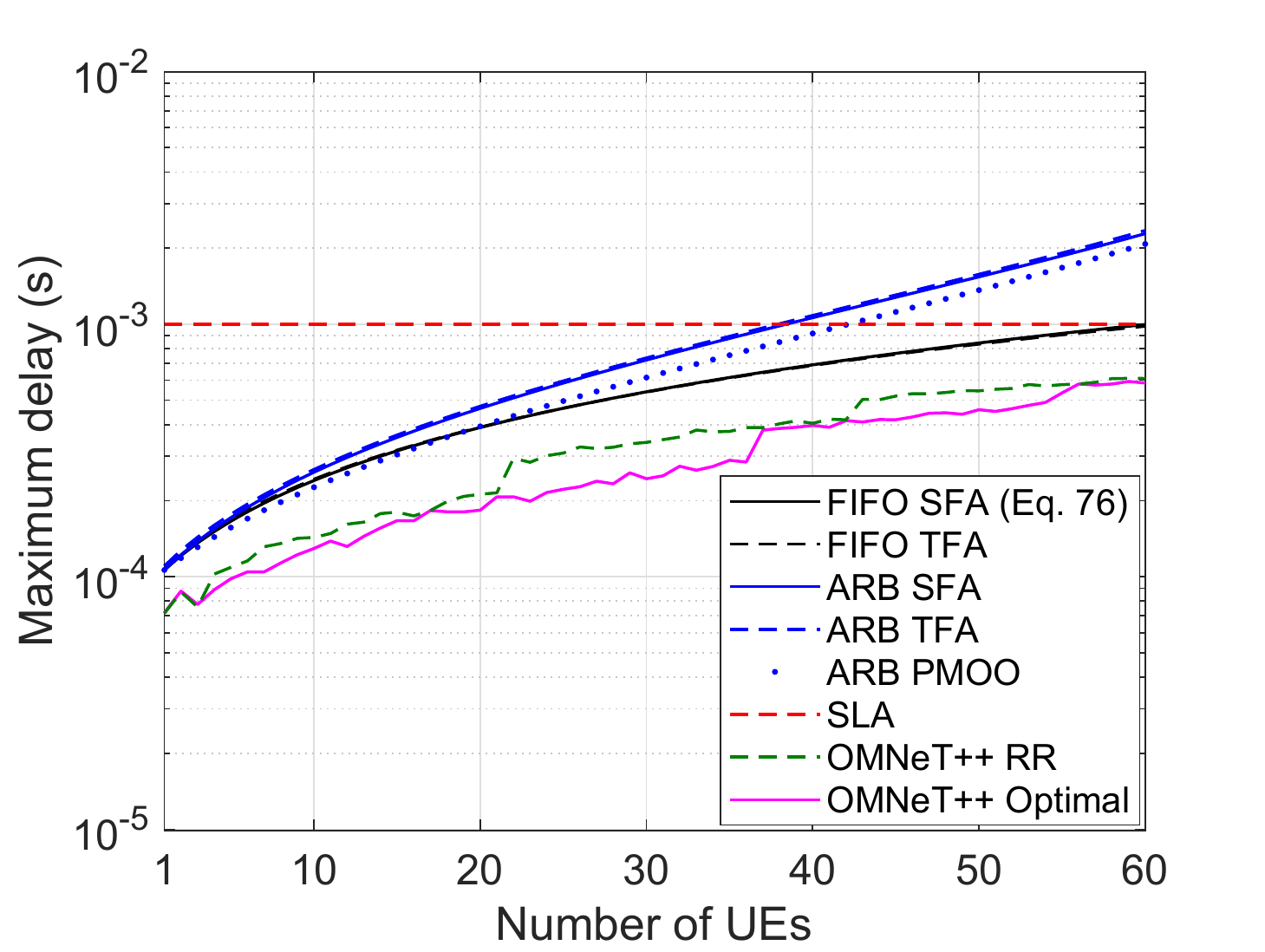}
    \caption{Maximum delay per number of UEs based on the optimal solution for vDU $u$ = 2.}
    \label{fig:UEs_latencia_CRAN_ES}
\end{figure}

In Fig. \ref{fig:UEs_latencia_CRAN_ES}, we show for vDU $u$ = 2 --- the vDU with split option $O_9$ admitting the largest number of UEs --- the maximum delay varying the number of UEs. We show in this figure the theoretical bounds provided by Network Calculus Theory using TFA and SFA analyses under FIFO and Arbitrary (ARB) multiplexing. We obtain these bounds using the optimal solution for bandwidth allocation at each transport node, routing path throughout the TN and functional split per vDU while varying the number of simultaneous UEs at vDU $u$ = 2 in the range starting from a single UE to the maximum number of admitted UEs (60 in the plot under analysis). The ARB multiplexing provides looser bounds than the FIFO multiplexing, mainly for a greater number of UEs in the network. This finding is explained by the fact that ARB multiplexing assumes no knowledge about the order of the packets to be transmitted, being a pessimist approach regarding the packet ordering in the queue, providing overestimated values for the end-to-end delay. In the figure under analysis, we see that using an ARB multiplexing based analysis instead of our proposal and fixing all other conditions, we would get approximately 40 UEs admitted to the network without SLA violation. Furthermore, we highlight that as more UEs arrive to the network, more pessimists are the bounds using ARB multiplexing. Contrasting SFA and TFA analyses, the obtained curves overlap each other, which is a direct consequence of the small burst size presented by URLLC applications. 


The PMOO analysis showed a slightly tighter bound than SFA and TFA analyses. All ARB multiplexing bounds violate the SLA. The chosen approach, presented in  (\ref{eq:de2e_app}), which is based on FIFO SFA analysis, offers the closest bound to the simulation results obtained from OMNeT++ and no SLA violation for the maximum number of UEs at the vDU under analysis.
Furthermore, the OMNeT++ RR curve, where RR stands for Round Robin scheduling, was obtained by fixing all optimal outputs except the transmission capacity allocation. The Round Robin scheduling shows higher delay values than those obtained using the optimal transmission capacity allocation (OMNeT++ Optimal), as we observe in Fig.~\ref{fig:UEs_latencia_CRAN_ES}. However, even for the RR scheduling, there is no SLA violation, thanks to all other optimal outputs used to obtain this result through simulation (optimal route, functional split, and number of admitted UEs).

On the one hand, we observe that the functional split deployment in the TN (midhaul) has a greater impact on the end-to-end delay than the bandwidth allocation per transport node (e.g., refer to the discrepant delay values observed in vDU $u$ = 1 and $u$ = 2 in Fig. \ref{fig:boxplot2}). On the other hand, in future real-time applications of our proposal, the bandwidth allocation may be performed in smaller time intervals and with higher frequency than modifications in functional split deployments with an important impact in the admission of UEs to the network (e.g., consider a stringent delay requirement of 0.5 ms instead of 1 ms in a simulation with OMNeT++, by fixing all other decision variables, we observe in Fig. \ref{fig:UEs_latencia_CRAN_ES} that the optimal bandwidth allocation would allow the admission of 55 UEs in vDU $u$ = 2 without SLA violation instead of 42 UEs using classical RR scheduling). 

\section{Related Work}\label{sec:RelatedWorks}

\begin{table*}[htb]
\begin{threeparttable}
\caption{Related Work}
\label{tab:relatedworks}
	\centering

	\begin{tabular}{|c|c|c|c|c|c|c|c|c|c|}
		\hline
		 \multirow{2}{*}{\textbf{Article}} & \multirow{2}{*}{\textbf{Slicing}} &\multirow{2}{*}{\textbf{Queueing}}& \multicolumn{2}{c|}{\textbf{Splitting}}  &  \multicolumn{2}{c|}{\textbf{Optimization}} & \multicolumn{3}{c|}{\textbf{Decision}}\\  \cline{4-10}
   &  & & \textbf{Flexible} & \textbf{Nodes} & \textbf{Objective}  & \textbf{Opt. Sol.} & \textbf{Bandwidth} & \textbf{Split Choice} & \textbf{Routing}\\ 
            \hline
            \hline
        \rowcolor{Gray}
  		WizHaul \cite{garcia2018wizhaul}  &$\times$  &  $\times$ & $\checkmark$ &RU-CU &  MAX CD  &$\times$ & $\times$& $\checkmark$ & $\checkmark$\\
		\hline
    FluidRAN \cite{garcia2018fluidran} &$\times$  &  $\times$  & $\checkmark$ & RU-CU&   MIN COST  & $\checkmark$ & $\times$& $\checkmark$ &$\checkmark$\\
		\hline
            \rowcolor{Gray}
      	PlaceRAN \cite{morais2022placeran} &$\times$  &  $\times$  &$\checkmark$& RU-DU-CU&   MIN DRC  &$\checkmark$& $\times$& $\checkmark$ &$\checkmark$\\
       \hline
       Molner et. al \cite{molner2019optimization} &$\times$  &  $\times$ & $\checkmark$ &DU-CU &  MAX DU-CU\tnote{1}   &$\times$& $\times$& $\checkmark$ &$\checkmark$\\
       \hline
            \rowcolor{Gray}
SlicedRAN \cite{ojaghi2019sliced} & $\checkmark$ &  $\times$  & $\checkmark$& RRH-CU &   MAX CD\tnote{2} & $-$& $\times$& $\checkmark$ &$\checkmark$\\
            \hline
            Ojaghi et. al  \cite{ojaghi2022impact} & $\checkmark$ &  $\times$  & $\checkmark$& DU-CU&   $\times$ & $\times$ & $\times$& $\checkmark$ &$\times$\\
            \hline
                      \rowcolor{Gray}
            SLICE-HPSO \cite{matoussi2020user} & $\checkmark$ & $\times$  &$\checkmark$ & DU-CU &   MAX RATE-COST\tnote{3} & $\times$ & $\checkmark$& $\checkmark$ &$-$\\
            \hline 
            Han et. al \cite{han2019utility} & $\checkmark$ & $\checkmark$\tnote{5}  & $\times$& $\times$ &  $-$ & $-$ & $-$& $\times$ &$\times$\\
            \hline
                      \rowcolor{Gray}

            Motalleb et. al \cite{motalleb2022resource} & $\checkmark$ & $\checkmark$\tnote{5}  & $\times$\tnote{4}& RU-DU-CU &  MAX RATE & $\times$ & $\checkmark$& $\times$ &$\times$\\
            \hline
                            Adamuz et. al \cite{adamuz2022stochastic} & $\checkmark$ & $\checkmark$ & $\times$ & $\times$ &  $\times$ & $\times$ & $-$& $\times$ &$-$\\
            \hline 
                      \rowcolor{Gray}
             Our Proposal & $\checkmark$ & $\checkmark$  & $\checkmark$ & RU-DU-CU &  MAX PROFIT& $\checkmark$ & $\checkmark$& $\checkmark$ & $\checkmark$\\
            \hline
  
\end{tabular}
  \begin{tablenotes}
   \item[1] Maximization of the number of DU deployments while minimizing the number of supporting CUs.
   \item[2] Jointly optimize the CD and throughput.
   \item[3] Jointly maximize user throughput and minimize cost.
   \item[4] Fixed functional split following O-RAN specifications.
   \item[5] Based on average measures (e.g., mean delay).   
  \end{tablenotes}
 \end{threeparttable}
\end{table*}

Our work involves network slicing, functional splitting, routing, queue modeling, and optimization problem formulation. In this context, the related literature deals with part but not all these aspects as shown in Table \ref{tab:relatedworks}, where we classify the works listed in the first column in terms of their adopted approaches, checking whether they deal with Slicing and Queueing in the second and third columns, respectively. In the fourth column, we show whether the works deal with functional splitting, specifying whether they do it in a flexible manner and showing the nodes the disaggregated RAN is split into. In the fifth column, we organize the cited works in terms of their problem formulation, detailing the objective function and whether the authors obtained optimal solutions. In the last column, we check the outputs obtained from the algorithms and models presented in the literature, showing whether their schemes provide decisions about the amount of bandwidth for each slice, the split choice to be deployed at the RAN nodes, and the routing paths along the transport network. We use `$\checkmark$' and `$\times$' to represent adopted and non-adopted approaches, respectively, and `$-$' to identify undeclared or unclear information.

\textbf{Functional Split and Routing.} A lot of work has been published regarding RAN disaggregation and virtualization \cite{foukas2016flexran,chang2017flexcran,garcia2018fluidran,garcia2018wizhaul,foukas2017orion}. Among them, we may highlight those with a flexible functional split. In \cite{garcia2018wizhaul}, the authors propose to maximize the Centralization Degree (CD) of VNFs while optimizing routing in the fronthaul transport network, assuming a disaggregated RAN with RU and CU nodes. The proposal is named WizHaul and considers a range of options for a flexible RAN split. WizHaul uses network delay as one of the optimization constraints as well as a cost function for VNF placement. However, it does not consider queueing effects nor provide maximum delay assurance. Moreover, WizHaul does not consider the network slicing paradigm. 

FluidRAN \cite{garcia2018fluidran} works with flexible functional split and routing. FluidRAN provides a solution for an optimization-based problem of cost minimization for MNOs. The system model considers the CD of RAN VNFs and routing paths between RU and CU. FluidRAN assumes that processing data at the CU is a lower-cost solution in comparison to RU. Therefore, the obtained results favor C-RAN solutions. In contrast, the authors analyze scenarios where RAN has to support \ac{MEC} services benefiting a decentralized VNF choice, creating a more complex layout where a solution is not trivial and relies on a trade-off decision, allowing operators to adjust the best scheme taking into account several underlying network conditions.

In a similar manner to WizHaul \cite{garcia2018wizhaul} and FluidRAN \cite{garcia2018fluidran}, PlaceRAN \cite{morais2022placeran} offers valuable contributions to the state-of-the-art in disaggregated RAN for 5G and B5G networks, addressing the problem of VNF placement for network planning purposes. PlaceRAN minimizes the number of \acp{CR} while maximizing the aggregation of VNFs, coining the term \ac{DRC}. The authors found an optimal solution that minimizes the sum of DRC values, delivering the best VNF placement and traffic routing for disaggregated RANs while reducing \acp{CR} and increasing the aggregation of VNFs.

The work in \cite{molner2019optimization} also addresses routing on the transport network and resource placement, considering disaggregated RAN scenarios, providing a heuristic to solve the formulated problem based on maximizing the number of DU deployments while minimizing the number of supporting CUs. Other works rely on \ac{AI}/\ac{ML} algorithms to jointly tackle routing and VNF placement \cite{he2023leveraging, xiao2019nfvdeep, quang2019multi, tong2020vnf}. None of the previously mentioned works considered the network slicing paradigm, a key feature for 5G networks. The following works took into account network slicing jointly with RAN disaggregation.


\textbf{Network Slicing and Functional Split}. 
In \cite{ojaghi2019sliced}, the authors present SlicedRAN, a framework for network slicing along functional splitting.  SlicedRAN proposes an optimization-based solution dealing with the trade-off between the following concerns: CD, computation cost, and throughput. However, assessment of the proposal viability is challenging since the work presents a preliminary performance evaluation, lacking comprehensive simulations and analyses of additional slicing use cases and comparisons with literature benchmarks.

Recently, the authors in \cite{ojaghi2022impact} discuss, shortly and as an overview, the impact of RAN \textit{densification} using a scenario with network slicing and functional split between DU and CU. The authors evaluate multiple slices per DU and propose the most suitable match between slice and functional split. The authors assume each slice has its appropriate split, and RUs are always co-localized with DU. Open questions not addressed in \cite{ojaghi2022impact} are: how to deploy multiple NSIs on a shared DU (with a single functional split); and how a single NSI can run on top of an infrastructure with multiple functional splits.

In \cite{matoussi2020user}, the authors present a user-centric approach, instead of a centralized one, to address the slice embedding problem in a disaggregated RAN with a flexible functional split. The authors formulate the problem using optimization to maximize user throughput and minimize cost. The proposal tackles radio, computational and link resources based on user, functional split, and slice requirements. The problem is solved using a \ac{PSO}-based heuristic.

None of the previously mentioned works considered queue modeling, even though queueing delay can critically impact time-sensitive applications such as those foreseen for the URLLC use case. Queue modeling becomes paramount in sliced and disaggregated RAN since the former implies resource sharing in multi-queueing scenarios and the latter implies multiple nodes --- and consequently multiple buffers --- along the transport network. The following works consider network slicing jointly with queue modeling.

\textbf{Network Slicing and Queue Modeling}. Using an MNO's perspective, the authors in \cite{han2019utility} propose an admission control solution based on a multi-queue-based controller that accounts for tenants' requests to network slices. For such a purpose, the authors use an $M/M/1$ queueing system and the analytical features of Little’s Formula, Steady Queue State Probability, and Waiting Time Distribution to describe the statistical tenant behaviors, optimizing the utility-based control for multiple and heterogeneous tenant requests. While the MNO leases infrastructure resources to tenants, the slicing use cases are not discussed, nor are the impacts of packet queueing and path routing along the transport network since this work covers only the MNO perspective.

In \cite{motalleb2022resource}, the authors deal with network slicing and VNF activation, considering the downlink of an O-RAN scenario with a fixed two-tier functional split. The main goal is resource allocation --- maximizing throughput --- concerning capacity and end-to-end delay constraints. The authors rely on the $M/M/1$ queueing theory for modeling transmission and processing delays, obtaining average measures of these metrics. The URLLC use case demands guarantees of no violations incompatible with average measures, while this approach provides simple and tractable equations to the problem statement. 

In \cite{adamuz2022stochastic}, the authors propose a RAN slicing planning strategy for MNOs using \ac{SNC} based queue modeling. The proposal concerns radio resource allocation while satisfying a stringent delay violation probability for the URLLC use case. In contrast with the \ac{DNC} approach, the \ac{SNC} framework allows small violations of the delay budget. Despite being attractive in relaxing the deterministic assumptions, such an approach is inappropriate for URLLC use cases demanding non-violation guarantees. Moreover, the authors do not tackle RAN functional split or routing.

Other works have presented similar approaches, dealing jointly with network slicing and queue modeling based on average measures, as in  \cite{tang2018queue}, or relying on \ac{QSI} to make decisions regarding resource allocation for different slices in a V2X application, as in \cite{khan2021network}, or allowing probabilistic violations for delay constraints in the context of haptic communications, as in \cite{aijaz2017mathsf}.
These works do not consider RAN disaggregation, which critically impacts delay budgets, nor routing in the transport network. Moreover, these works lack guarantee offers for critical URLLC applications. We have put it all together in our study, dealing jointly with network slicing, disaggregated RAN, queue modeling, and routing. Furthermore, we addressed this joint issue using an optimization problem formulation with an optimal solution. Our work is comprehensive, mathematically tractable, and in compliance with the latest evolving technologies in 5G network slicing and RAN disaggregation.

\section{Conclusions and Future Work}\label{sec:conclusions}

In this work, we have shown that differently from the full-stack RAN deployment, the disaggregated RAN implementation imposes challenges to the resource allocation for NSIs across the transport network. The main reason is the distinct disaggregated RAN nodes associated with various functional splits, leading to diverse packet overheads transmitted across the transport network and consequently providing different end-to-end delays for packets of the same NSI but transmitted to UEs associated with distinct RUs. This issue is paramount for delay-sensitive applications, such as those of the 3GPP URLLC use cases. 
To deal with this issue, we have formulated a latency-constrained optimization problem to maximize MNO's profit with optimal solutions to the transmission resource share allocated to each NSI at each transport node, VNF placement to each disaggregated RAN node, traffic routing throughout the transport network, and admission control of the number of UEs per RU in the network. We have leveraged the worst-case analysis provided by the deterministic network calculus theory, deriving a closed-form equation to estimate an end-to-end delay bound. 
Our proposal is tailored to offer guarantees of no violations of the delay requirements of the applications used by all URLLC UEs admitted to the network. The simulations that were carried out considering a 5G NG-RAN system scenario revealed that the proposed approach outperforms other schemes, 
highlighting the applicability of flexible functional splits to sliced and disaggregated RANs. We argue, corroborated by our results, that a one-size-fits-all solution based on a fixed functional split cannot efficiently handle the heterogeneity of a sliced, disaggregated, and shared RAN. The flexible approach designed by our proposal supports multiple slice instantiation over a shared infrastructure while reducing operational costs, increasing revenues obtained through URLLC applications, and, consequently, maximizing MNO's profit.

For future work, we aim to implement our solution in a real-world scenario. Specifically, we will deploy a testbed considering sliced and disaggregated RANs using Intel Tofino switches. We intend to use the optimal solution obtained through our formulation as input to this testbed scenario, leveraging the data plane programmability of Intel Tofino switches using the P4 language. We will measure, using telemetry techniques, the one-way end-to-end delay for comparison with that obtained from our theoretical approach. This comparison between theory and practice will be valuable for understanding the plausibility of the assumptions made in this work and assessing the accuracy of the delay bounds with actual equipment. Furthermore, by using programmable switches, we will be able to propose and evaluate online approaches considering real-world limitations.

\section*{Acknowledgement}
\addcontentsline{toc}{section}{Acknowledgement}
This work was conducted with partial financial support from the National Council for Scientific and Technological Development (CNPq) under Grant number 405111/2021-5 and 130555/2019-3 and from the Coordination for the Improvement of Higher Education Personnel (CAPES) - Finance Code 001, Brazil. Moreover, RNP partially supported the work, with resources from MCTIC, Grant number 01245.010604/2020-14, under the 6G Mobile Communications Systems project. This work was also supported by MCTIC/CGI.br/São Paulo Research Foundation (FAPESP) through grant number 2022/05397-5 and through projects Slicing Future Internet Infrastructures (SFI2) – grant number 2018/23097-3, Smart 5G Core And MUltiRAn Integration (SAMURAI) – grant number 2020/05127-2, Programmable Future Internet for Secure Software Architectures (PROFISSA) - grant number 2021/08211-7, Programmability, ORchestration and VIRtualization of 5G Networks (PORVIR-5G) - grant number 2020/05182-3.


\printacronyms[name=List of Acronyms,sort=true,template=supertabular]

\bibliographystyle{IEEEtran}
\bibliography{refs}


\end{document}